\RequirePackage[mathlines]{lineno} % Display line numbers
\documentclass[a4paper,11pt]{article}
\usepackage{jheppub} % for details on the 
\usepackage[T1]{fontenc} % if needed
\emailAdd{besiii-publications@ihep.ac.cn}
\usepackage{threeparttable}%This package is used for tablenotes
\usepackage{multirow}
\usepackage{subfigure}
\usepackage{float}
\usepackage{ulem}
\usepackage{color}
\usepackage[thicklines]{cancel}
\usepackage{tabularx} % for 'tabularx' env. and 'X' col. type
\usepackage{ragged2e} % for \RaggedRight macro
\usepackage{booktabs} % for \toprule, \midrule etc macros
\newcolumntype{L}{>{\RaggedRight\hangafter=1\hangindent=0em}X}
\usepackage[compat=1.0.0]{tikz-feynman}
\usepackage{feynmp-auto}
\usepackage{siunitx}
\usepackage[utf8]{inputenc}
\usepackage{orcidlink}
\DeclareUnicodeCharacter{2212}{-}
\DeclareUnicodeCharacter{0327}{\c}

\let\oldequation\equation
\let\oldendequation\endequation
\renewenvironment{equation}
{\linenomathNonumbers\oldequation}
{\oldendequation\endlinenomath}
\def \gev  {\mbox{GeV}}

\def \mev  {\mbox{MeV}}

\begin{document}
	
	\title{\boldmath Measurement of branching fractions of $\Lambda_{c}^{+}$ decays to $\Sigma^{+} \eta$ and $\Sigma^{+} \eta'$}
	
	\collaboration{The BESIII collaboration}
	
	\abstract{
		By analyzing $e^+e^-$ collision data taken at center-of-mass energies
		$\sqrt{s}$ between 4.600  and 4.699 $\gev$ with the BESIII detector at the BEPCII collider, corresponding to an integrated luminosity of  $\rm 4.5~fb^{-1}$, we study the hadronic decays $\Lambda_{c}^{+} \rightarrow \Sigma^{+} \eta$ and $\Lambda_{c}^{+} \rightarrow \Sigma^{+} \eta^{\prime}$ using the single-tag method. The branching fraction ratio of $\Lambda_{c}^+ \rightarrow \Sigma^+ \eta$ relative to $\Lambda_{c}^+ \rightarrow \Sigma^+ \pi^0$ is determined to be $0.305 \pm 0.046_{\rm stat.} \pm 0.007_{\rm syst.}$,
		and that of $\Lambda_{c}^+ \rightarrow \Sigma^+ \eta'$ relative to $\Lambda_{c}^+ \rightarrow \Sigma^+ \omega $ is $0.336 \pm 0.094_{\rm stat.} \pm 0.037_{\rm syst.}$. The ratio of 
		$\frac{\mathcal{B}\left(\Lambda_{c}^{+} \rightarrow \Sigma^{+} \eta'\right)}{\mathcal{B}\left(\Lambda_{c}^{+} \rightarrow \Sigma^{+} \eta\right)}
		$ is determined to be 
		$1.73 \pm 0.22_{\rm stat.} \pm 0.16_{\rm syst.}$.
		These results enrich our knowledge of charmed baryon decays. 
	}
	
	\keywords{ BESIII, charmed baryon, Cabibbo-Favored decay}
	
	\arxivnumber{2505.18004}
	
	\maketitle
	\flushbottom
	
	\section{INTRODUCTION}
	\label{sec:introduction}
	
	Nonleptonic decays of charmed baryons offer excellent opportunities for testing different theoretical approaches to describe the complicated dynamics of heavy-light baryons, including the current algebra approach~\cite{1994TUppal}, the factorization scheme, the pole model technique~\cite{1992QPXu,1999Sharma,2020Zou,1994Zenczykowski}, the relativistic quark model~\cite{1992Korner,1998Ivanov}, the $SU(3)$ flavor symmetry~\cite{2020Zhao,2019Geng,Cheng_2025oyr}, and the quark-diagram scheme~\cite{PhysRevD.54.2132,PhysRevLett.56.1655}. Contrary to the significant progress made in the studies of heavy meson decays, the progress in both theoretical and experimental studies of heavy baryon decays has been relatively slow.
	Since the first observation of the ground state $\Lambda_{c}^{+}$ at the Mark II experiment in $1979$~\cite{1980GSAbrams}, many decays are still unknown or lack measurement accuracy~\cite{pdg2022}.
	
	\begin{figure}[htb]
		\centering
		\subfigure[Internal W-emission.]{\includegraphics[width=0.4\textwidth]{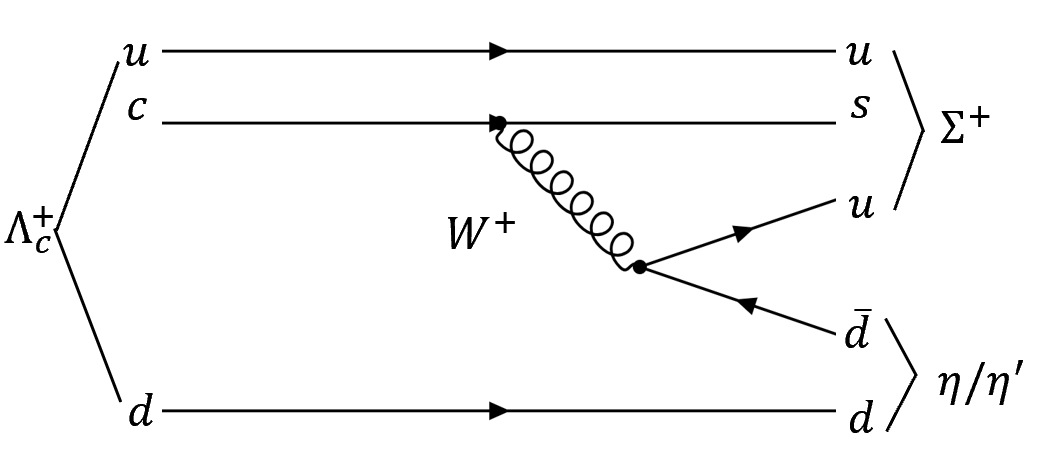}}
		\subfigure[W-exchange.]{\includegraphics[width=0.4\textwidth]{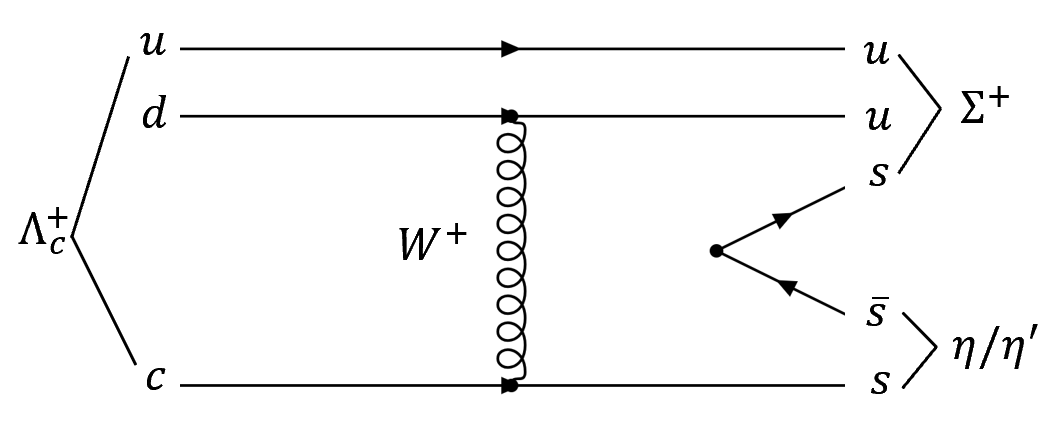}} \\
		\caption{Feynman diagrams of the $\Lambda_{c}^+ \rightarrow \Sigma^+ \eta/\eta'$ decay.}
		\label{fig:feynman}
	\end{figure}

	The two-body Cabibbo-favored (CF) decay of the $\Lambda_{c}^{+}$ to an octet baryon $B$ and a pseudoscalar meson $P$, $\Lambda_{c}^{+} \rightarrow B\left(\frac{1}{2}^{+}\right) P$,
	is one of the simplest hadronic channels to be treated theoretically~\cite{1993HYCheng}.
	However, existing model calculations still yield different  branching fractions~(BFs) predictions, precise BF measurements are essential for calibrating and discriminating among the various theoretical approaches.
	% But different theoretical models give different results, and the measurement of the branching fractions (BFs) is important  to calibrate the different theoretical approaches. 
	The $\mathrm{CF}$ decays $\Lambda_{c}^{+} \rightarrow \Sigma^{+} \eta$ and $\Lambda_{c}^{+} \rightarrow \Sigma^{+} \eta^{\prime}$, which proceed entirely through nonfactorizable internal $W$-emission and $W$-exchange diagrams as shown in Figure~\ref{fig:feynman}, are particularly interesting. Unlike charmed meson decays, they are not affected by color or helicity suppression and have therefore relatively sizable BFs. Theoretical predictions of the nonfactorizable diagrams are not very consistent, resulting in large variations of the predicted BFs, \textit{e.g.}, $\mathcal{B}\left(\Lambda_{c}^{+} \rightarrow \Sigma^{+} \eta\right)=(0.11-0.90) \%$ and $\mathcal{B}\left(\Lambda_{c}^{+} \rightarrow\right.$ $\left.\Sigma^{+} \eta^{\prime}\right)=(0.10-1.44) \%$~\cite{1999Sharma,1992Korner,1994Zenczykowski,1998Ivanov,2019Geng,2020Zou,2020Zhao, Cheng_2025oyr}. 
	Several experimental measurements have been performed on these BFs.
	CLEO reported the evidence of the decay $\Lambda_{c}^{+} \rightarrow \Sigma^{+} \eta$ with a BF of $(0.70 \pm 0.23) \%$~\cite{1995RAmmar}.
	Belle has measured  $\mathcal{B}\left(\Lambda_{c}^{+} \rightarrow \Sigma^{+} \eta\right)=(3.14\pm0.35\pm0.11\pm0.25)\times10^{-3}$ and $\mathcal{B}\left(\Lambda_{c}^{+} \rightarrow \Sigma^{+} \eta'\right)=(4.16\pm0.75\pm0.21\pm0.33)\times10^{-3}$~\cite{2023Belle}.
	BESIII has measured  $\mathcal{B}(\Lambda_{c}^{+} \rightarrow \Sigma^{+} \eta)=(0.41 \pm 0.19 \pm 0.05) \%$ 
	and $\mathcal{B}(\Lambda_{c}^{+} \rightarrow \Sigma^{+} \eta')=(1.34 \pm 0.53 \pm 0.19)\%$
	using $e^+e^-$ collision data taken at the center-of-mass system (CMS) energy $\sqrt s=4.600$ GeV~\cite{BESIII:2018cdl} corresponding to an integrated luminosity of $\rm 567~pb^{-1}$;
	an improvement of the measurements of these decays is expected  with the larger data sets newly accumulated by BESIII. 
	Furthermore, the ratio ${\mathcal{B}\left(\Lambda_{c}^{+} \rightarrow \Sigma^{+} \eta'\right)}/{\mathcal{B}\left(\Lambda_{c}^{+} \rightarrow \Sigma^{+} \eta\right)}
	= 1.34 \pm 0.28 \pm 0.06$ has been measured by Belle~\cite{2023Belle}, consistent within $2\sigma$ with the theoretical predictions from Refs.~\cite{1999Sharma,1994Zenczykowski}, but consistent with  Refs.~\cite{Cheng_2025oyr} closely, where $\sigma$ is standard deviation.  
	BESIII reported ${\mathcal{B}\left(\Lambda_{c}^{+} \rightarrow \Sigma^{+} \eta'\right)}/{\mathcal{B}\left(\Lambda_{c}^{+} \rightarrow \Sigma^{+} \eta\right)} = 3.5 \pm 2.1 \pm 0.4$~\cite{BESIII:2018cdl}, consistent with the Belle result, albeit with large uncertainty.  
	Further experimental studies of these two decays
	are essential for testing different theoretical models and gaining a better understanding of the $\Lambda_{c}^{+}$ CF decays.
	
	In this work, we present a measurement of the 
	branching ratios ${\mathcal{B}\left(\Lambda_{c}^{+}\rightarrow\Sigma^{+} \eta\right)}/{\mathcal{B}\left(\Lambda_{c}^{+} \rightarrow \Sigma^{+} \pi^0\right)}$
	and ${\mathcal{B}\left(\Lambda_c^+ \rightarrow \Sigma^+ \eta'\right)}/{\mathcal{B}\left(\Lambda_c^+ \rightarrow \Sigma^+ \omega\right)}$
	by analyzing $e^+e^-$ collision data  taken at $\sqrt{s}=4.600$, 4.612, 4.628, 4.641, 4.661, 4.682, and 4.699~GeV~\cite{BESIII:2022ulv} with the BESIII detector at the BEPCII collider, corresponding to an integrated luminosity of 4.5 fb$^{-1}$, as detailed in Table~\ref{tab:sum}. Throughout this paper, charge-conjugate modes are implicitly assumed. In Sect.~\ref{sec:detector}, the BESIII detector and the data samples are described. The event selection is introduced in Sect.~\ref{sec:analysis}. The measurement of the BFs is presented in Sect.~\ref{sec:bf}. The systematic uncertainties are discussed in Sect.~\ref{sec:systematic}. Finally, Sect.~\ref{sec:summary} summarizes the results. 
	%%%%%%%%%%%%%
	\begin{table}[!htbp]
		\footnotesize
		\caption{The CMS energies  $E_{\rm CMS}$ and the integrated luminosities $\int\!L\,dt$ at different energy points. The first and second uncertainties are statistical and systematic, respectively.}
		\begin{center}
			\begin{tabular}{cSS}
				\hline
				\hline
				{Sample}  & {$E_{\rm CMS}$ ($\mev$)}  & {$\int\!L\,dt$ (pb$^{-1}$)}\\
				\hline
				4.600& 4599.53\pm0.07\pm0.74&586.90\pm0.10\pm3.90\\
				4.612& 4611.86\pm0.12\pm0.30&103.65\pm0.05\pm0.55\\
				4.628& 4628.00\pm0.06\pm0.32&521.53\pm0.11\pm2.76\\
				4.641& 4640.91\pm0.06\pm0.38&551.65\pm0.12\pm2.92\\
				4.661& 4661.24\pm0.06\pm0.29&529.43\pm0.12\pm2.81\\
				4.682& 4681.92\pm0.08\pm0.29&1667.39\pm0.21\pm8.84\\
				4.699& 4698.82\pm0.10\pm0.36&535.54\pm0.12\pm2.84\\
				\hline 
				\hline
			\end{tabular}
		\end{center}
		\label{tab:sum}
	\end{table}
	
	\section{BESIII DETECTOR AND MONTE CARLO SIMULATION}
	\label{sec:detector}
	The BESIII detector~\cite{Ablikim:2009aa} records symmetric $e^+e^-$ collisions 
	provided by the BEPCII storage ring~\cite{Yu:IPAC2016-TUYA01}
	in the center-of-mass energy range from 1.85 to 4.95~GeV,
	with a peak luminosity of $1.1 \times 10^{33}\;\text{cm}^{-2}\text{s}^{-1}$ 
	achieved at $\sqrt{s} = 3.773\;\text{GeV}$. 
	BESIII has collected large data samples in this energy region~\cite{Ablikim:2019hff}. The cylindrical core of the BESIII detector covers 93\% of the full solid angle and consists of a helium-based
	multilayer drift chamber~(MDC), a time-of-flight system~(TOF), and a CsI(Tl) electromagnetic calorimeter~(EMC),
	which are all enclosed in a superconducting solenoidal magnet
	providing a 1.0~T magnetic field. The solenoid is supported by an
	octagonal flux-return yoke with resistive plate counter muon
	identification modules interleaved with the steel. 
	
	The charged-particle momentum resolution at $1~{\rm GeV}/c$ is
	$0.5\%$, and the specific energy loss ($dE/dx$) resolution is $6\%$ for electrons
	from Bhabha scattering. The EMC measures photon energies with a
	resolution of $2.5\%$ ($5\%$) at $1$~GeV in the barrel (end cap)
	region. The time resolution in the TOF barrel region is 68~ps, while
	that in the end cap region is 110~ps. The end cap TOF
	system was upgraded in 2015 using multi-gap resistive plate chamber
	technology, providing a time resolution of
	60~ps~\cite{etof,etof2,etof3}.  
	About 87\% of the data used in this analysis benefits from this upgrade.  
	
	Monte Carlo (MC) simulated data samples produced with the {\sc
		geant4}-based~\cite{geant4} software package {\sc boost}~\cite{bes:boost}, which
	includes the geometric and material description of the BESIII detector~\cite{geo2,detvis} and the detector responses, are used to determine detection efficiencies
	and to estimate backgrounds. The simulation models the beam
	energy spread and initial state radiation (ISR) in the $e^+e^-$
	annihilations with the generator {\sc
		kkmc}~\cite{ref:kkmc,ref:kkmc2}. 
	The inclusive MC sample, which is about 40 times the datasets,  
	includes the production of open charm
	processes, the ISR production of vector charmonium(-like) states, and the continuum processes.
	All particle decays are generated with {\sc
		evtgen}~\cite{ref:evtgen, PingRong_Gang_2008} using BFs either taken from the Particle Data Group~(PDG)~\cite{pdg2022}, when available, or otherwise estimated with {\sc lundcharm}~\cite{ref:lundcharm, ref:lundcharm2}. Final state radiation
	from charged final state particles is incorporated using the {\sc
		photos} package~\cite{photos}.
	For the MC production of the $e^+e^-\rightarrow \Lambda_c^+\overline{\Lambda}_c^-$ events, the cross section line-shape from BESIII measurements is taken into account.   
	 All the final tracks and photons are propagated through the GEANT4-based detector simulation framework.
	 
%	 passed into the detector simulation package.  
	
	\section{ANALYSIS METHOD AND EVENT SELECTION}
	\label{sec:analysis}
	\hspace{1.5em} 
	
	A single-tag method is applied in this work, which means only $\Lambda_c^+$ or $\bar{\Lambda}_{c}^{-}$ is reconstructed in an event. Both the signal and the reference $\Lambda_c^+$ decays are fully reconstructed.  
	Charged tracks detected in the MDC are required to be within a polar angle ($\theta$) range of $|\cos\theta|<0.93$, where $\theta$ is the polar angle defined with respect to the $z$ axis, which is the symmetry axis of the MDC. The distance of closest approach to the interaction point (IP) must be less than 10\,cm along the $z$ axis and less than 1\,cm in the transverse plane.  
	
	Particle identification~(PID) for charged tracks combines measurements of the specific ionization energy loss in the MDC~(d$E$/d$x$) and the flight time measured in the TOF to form likelihoods $\mathcal{L}(h)~(h=p,K,\pi)$ for each hadron $h$ hypothesis.
	Tracks are identified as protons when the proton hypothesis has the largest likelihood ($\mathcal{L}(p)>\mathcal{L}(K)$ and $\mathcal{L}(p)>\mathcal{L}(\pi)$), while charged kaons and pions are identified by comparing the likelihoods for the kaon and pion hypotheses, $\mathcal{L}(K)>\mathcal{L}(\pi)$ and $\mathcal{L}(\pi)>\mathcal{L}(K)$, respectively.
	
	Photon candidates are reconstructed using showers in the EMC.  The deposited energy of each shower must be greater than 25~MeV in the barrel region ($|\cos \theta|< 0.80$) and greater than 50~MeV in the end cap region ($0.86 <|\cos \theta|< 0.92$). To exclude showers that originate from charged tracks, the angle subtended by the EMC shower and the position of the closest charged track at the EMC must be greater than 10 degrees as measured from the IP. To suppress electronic noise and showers unrelated to the event, the difference between the EMC time and the event start time is required to be within [0, 700]\,ns.
	
	Both $\eta$ and $\pi^0$ candidates are reconstructed by combinations of photon pairs. First,  photon pairs are considered as $\pi^{0}$ candidates, for which the reconstructed invariant mass $M(\gamma\gamma)$ is required to fall in the range  
	$0.115$~GeV/$c^{2}<M(\gamma\gamma)<0.150$~GeV/$c^{2}$. The remaining photon pairs are considered as $\eta$ candidates when $0.500$~GeV/$c^{2}<M(\gamma\gamma)<0.560$~GeV/$c^{2}$. 
	The $\Sigma^+$ candidates are reconstructed from  combinations of protons and $\pi^0$ with an invariant mass $1.174$~GeV/$c^2< M(p\pi^0)< 1.200$~GeV/$c^2$.
	The $\omega$ candidates are reconstructed from $\pi^+\pi^-\pi^0$ combinations with an invariant mass $0.760$~GeV/$c^2< M(\pi^+\pi^-\pi^0)< 0.800$~GeV/$c^2$.
	The $\eta'$ candidates are reconstructed from the $\pi^+\pi^-\eta$ combinations with an invariant mass $0.946$~GeV/$c^2< M(\pi^+\pi^-\eta)< 0.968$~GeV/$c^2$.
	For the $\Lambda_c^+ \rightarrow \Sigma^+ \eta$ decay, the $\eta \rightarrow \pi^+ \pi^- \pi^0$ decay is also used to reconstruct the $\eta$ meson, requiring $0.535$~GeV/$c^2< M(\pi^+\pi^-\pi^0)< 0.560$~GeV/$c^2$.
	All mass requirements on the candidates correspond to approximately $\pm3\sigma$ (standard deviations) around the individual nominal masses taken from the PDG ~\cite{pdg2022}. 
	Another requirement of $M(\pi^{0} \pi^{0}) \notin (0.440,0.520) ~\mathrm{GeV/}c^2$  is applied to veto background from $\Lambda_c^+ \to p K_S^0$ decays.
	
	To improve the energy resolution of photons from $\eta$ or $\pi^0$ decays, we constrain the two-photon invariant mass from $\eta, \pi^0$ decays to the nominal mass of $\eta, \pi^0$ given by PDG\cite{pdg2022}.
	To select the best combination among multiple $\Lambda_c$ candidates in each event,
	a kinematic fit constraining the invariant mass of the system recoiling against the reconstructed $\Lambda_{c}^{+}$ candidate to the nominal mass of the $\bar{\Lambda}_{c}^{-}$ is applied.  The combination with the smallest $\chi^2$ of this fit is retained.
	After that, we update the momenta of all final state particles for the further analysis.
	For the decays $\Lambda_{c}^+ \rightarrow \Sigma^+ \eta$ and $\Lambda_{c}^+ \rightarrow \Sigma^+ \pi^0$, we choose the combination with the minimum fit quality $\chi^2$ as the best candidate.  For the decay  $\Lambda_{c}^+ \rightarrow \Sigma^+ \eta'$, only the one with the minimum mass difference $\left|M\left(\pi^{+} \pi^{-} \eta\right)-M\left(\eta^{\prime}\right)\right|$ is accepted.
	Similarly, the combination with the invariant mass  $M\left(\pi^{+} \pi^{-} \pi^0\right) $
	closest to the $\omega$ mass
	is accepted for the reference decay $\Lambda_{c}^+ \rightarrow \Sigma^+ \omega$.

	To further suppress the combinatorial backgrounds in  $\Lambda_{c}^+ \rightarrow \Sigma^+ \eta$ and $\Lambda_{c}^+ \rightarrow \Sigma^+ \eta'$ channels, an anti-proton recoiling against the detected $\Lambda_c^+$  candidate is required, which is expected to originate from the $\bar{\Lambda}_c^-$ decay~\cite{BESIII:2018cdl}. The reference decays use the same method to reduce the systematic uncertainties.
	
	We  maximize a figure of merit (FOM) defined as $\frac{s}{\sqrt{s+b}}$, where $s$ is the signal yield with its associated BF given by PDG~\cite{pdg2022} and $b$ is the background yield, to optimize the $\chi^2$ cuts by using inclusive MC samples normalized to the data luminosity. 
	For the  decays $\Lambda_{c}^+ \rightarrow \Sigma^+ \eta$ and $\Lambda_{c}^+ \rightarrow \Sigma^+ \eta'$ we require  $\chi^2<17$ and $\chi^2<30$, respectively.
	
\begin{figure}[htbp]
		\centering
		\subfigure[$\Lambda_{c}^+ \rightarrow \Sigma^+ \eta(\gamma \gamma) $.]{\includegraphics[width=0.3\textwidth]{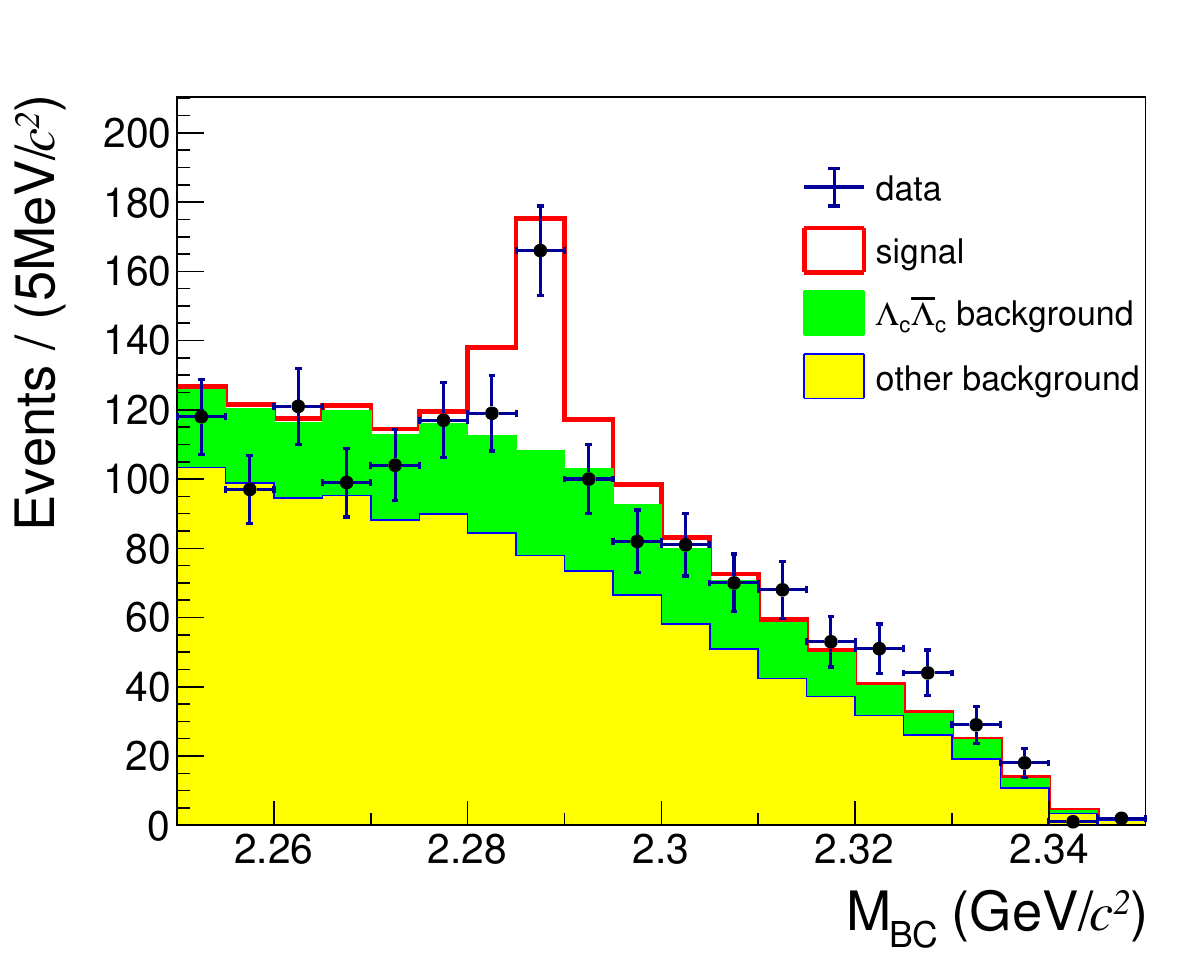}}
		\subfigure[$\Lambda_{c}^+ \rightarrow \Sigma^+ \eta(\pi^+\pi^-\pi^0) $.]{\includegraphics[width=0.3\textwidth]{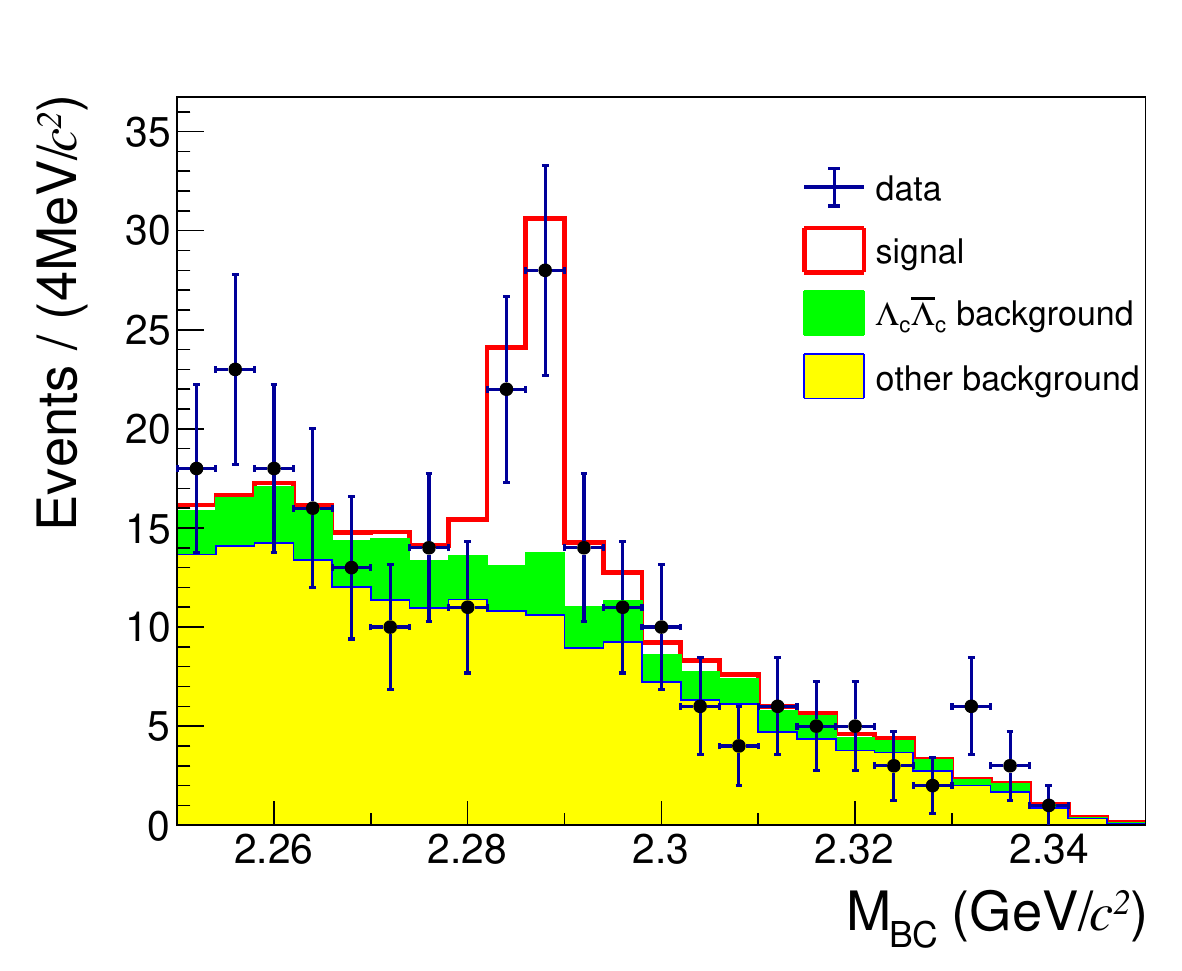}}
		\subfigure[$\Lambda_{c}^+ \rightarrow \Sigma^+ \eta'$]{\includegraphics[width=0.3\textwidth]{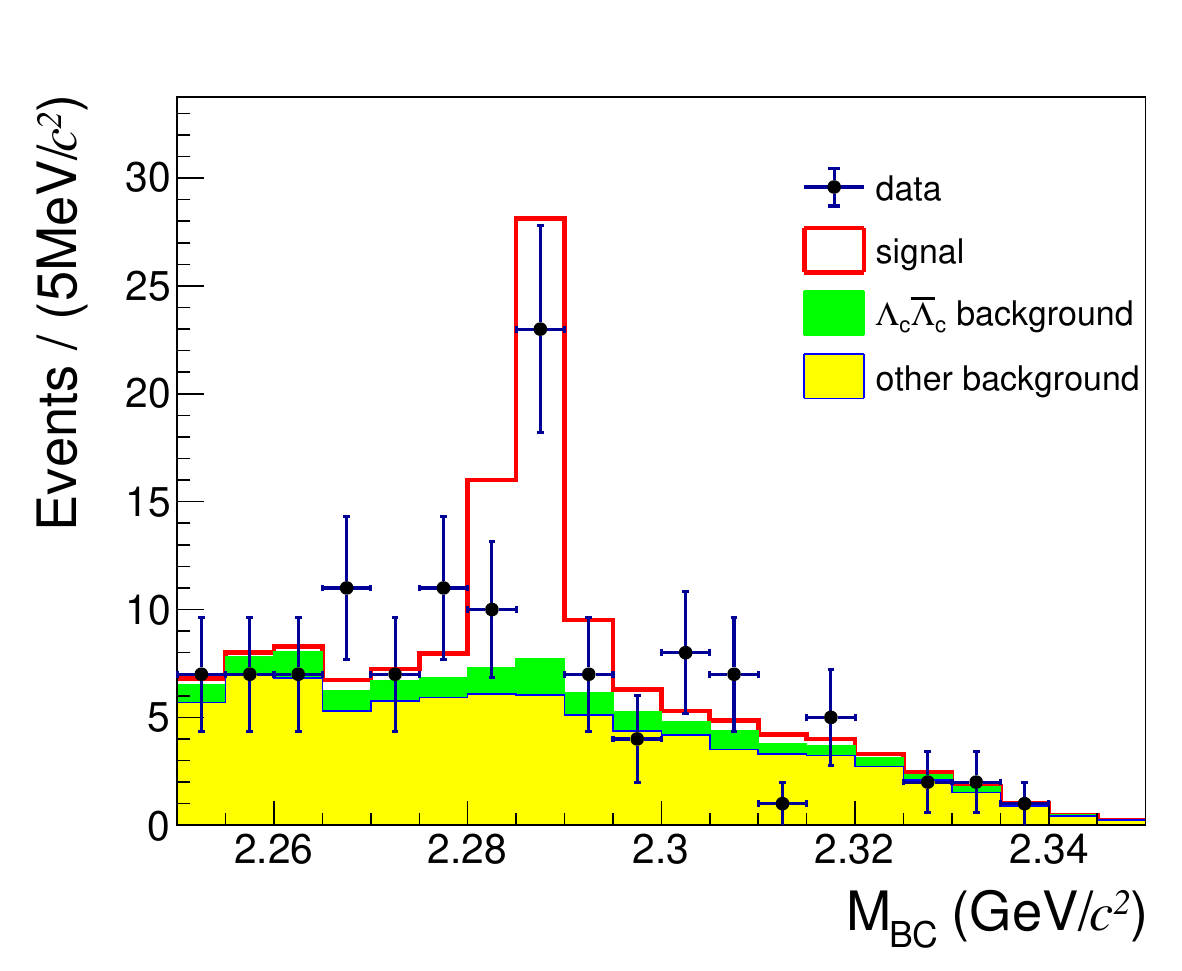}}\\
		\subfigure[$\Lambda_{c}^+ \rightarrow \Sigma^+ \pi^0$]{\includegraphics[width=0.3\textwidth]{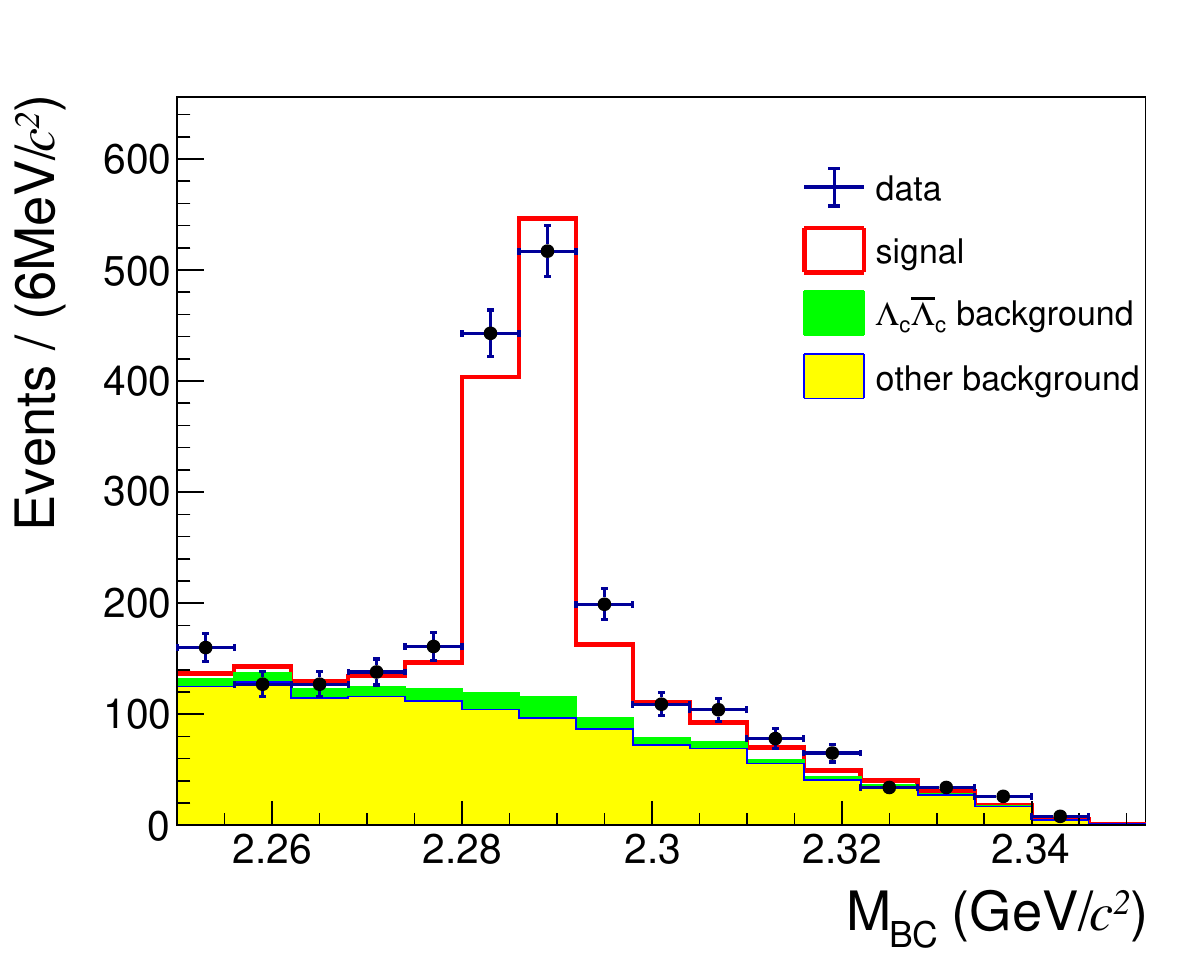}} 
		\subfigure[$\Lambda_{c}^+ \rightarrow \Sigma^+ \omega$]{\includegraphics[width=0.3\textwidth]{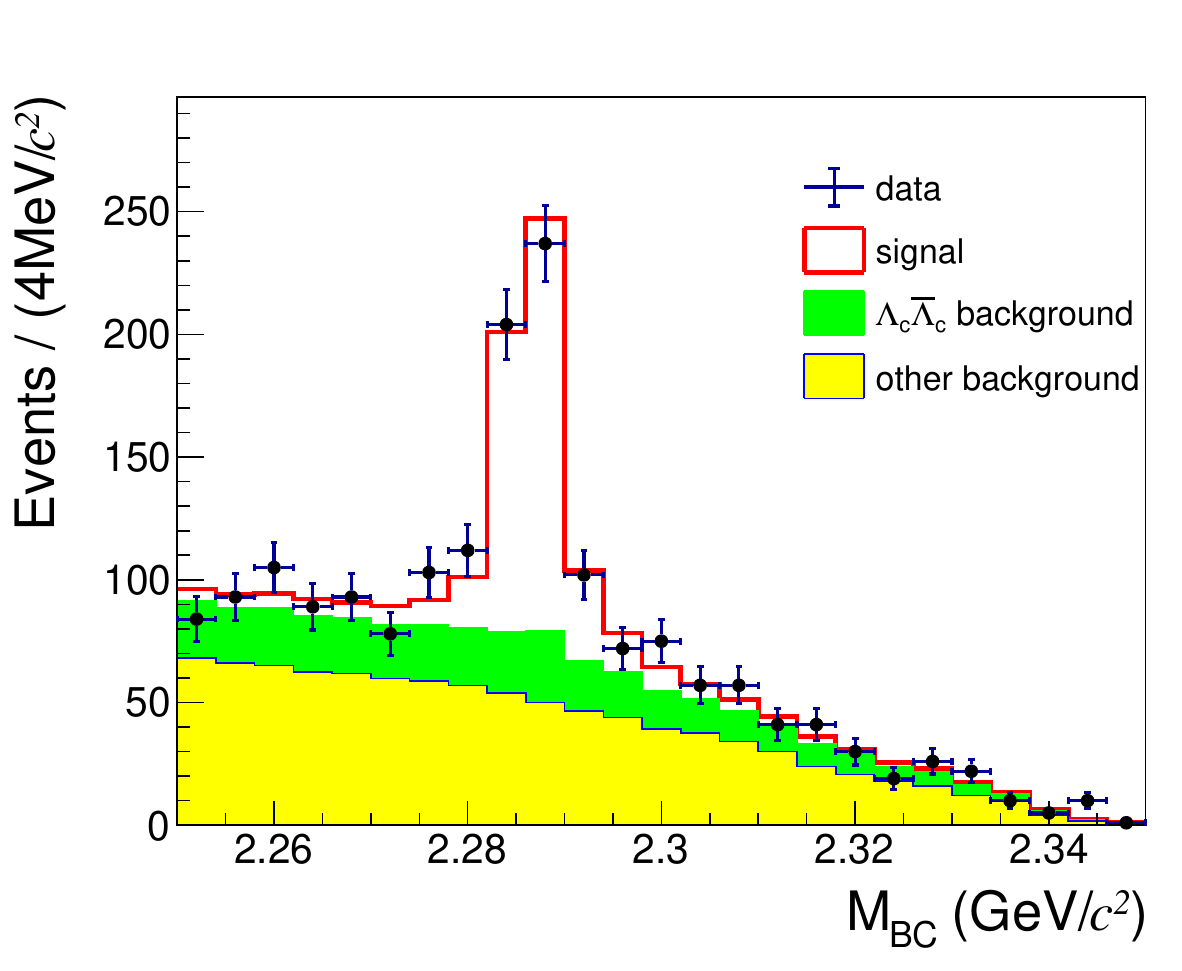}}\\
		
		\caption{Distributions of $M_{\rm BC}$ of accepted candidates for each decay, combined from all energy points. The MC samples have been scaled to that of data. The black dots with error bars are data combining all energy points, the red lines show the signal, the green histograms represent the $\Lambda_c\bar{\Lambda}_c$ backgrounds, and the yellow histograms represent the other backgounds.}
		\label{fig:checkbkg}
	\end{figure}

	The $\Lambda^+_c$ signal is identified using the beam constrained mass $M_{\rm BC} = \sqrt{E_{\rm beam}^2/c^4 - p^2/c^2}$, where $p$ is the measured $\Lambda_c^+$ momentum in the CMS of the $e^+e^-$ collision and $E_{\rm beam}$ is the beam energy.  After the event selection, there is no obvious peaking background in the $M_{\rm BC}$ distributions for each considered decay channel, as shown in Figure~\ref{fig:checkbkg}.
	
	\section{DETERMINATION OF THE BRANCHING FRACTIONS}
	\label{sec:bf}
	A simultaneous unbinned maximum likelihood fit on the $M_{\rm BC} $ distributions at the different energy points is performed to extract the signal yield.
	For the data sets taken at the  seven different CMS energies, the likelihood function is  written as
	\begin{equation}
		\mathcal{L}_{j}^{\text {total }} = \prod_{j} \mathcal{L}_{j},
	\end{equation}
	where the index $j$ denotes the energy point. At the $j-$th  energy point, the likelihood is written as
	\begin{equation}
		\mathcal{L}_{j}=\mathcal{L}_{j}^{\text {sig}} \cdot \mathcal{L}_{j}^{\text {ref}},
	\end{equation}
	where $\mathcal{L}_{j}^{\text {sig }} $ and $ \mathcal{L}_{j}^{\text {ref }} $ are the likelihood functions of the signal and the reference decays. Since they are similar to each other, 
	we omit the superscript ${\mathrm {sig}}$ or ${\mathrm {ref}}$ and 
	obtain the likelihood  function as
	\begin{equation}
		\mathcal{L}_{j}\left(N_{j}, \vec{\lambda}_{j}\right)=\frac{e^{-N_{j}} N_{j}^{N_{j}^{\mathrm{obs}}}}{N_{j}^{\mathrm{obs}} !} \prod_{j}^{N_{j}^{\mathrm{obs}}} \mathcal{P}_{j}\left({x}_{j} ; \vec{\lambda}_{j}\right).
		\label{equa:L_j}
	\end{equation}
	The $N_j^{\rm obs} $ is the observed yield of the signal or reference channel. 
	For the signal decays, the Probability Density Functions (PDFs), parameterized as the sum of signal and background PDFs, are written as 
	\begin{equation}
		\mathcal{P}_{j}\left({x}_{j} ; \vec{\lambda}_{j}\right)=\frac{N_{j}^{\mathrm{s}}}{N_{j}} \cdot\left[\mathcal{P}_{j}^{\mathrm{s}} \otimes \operatorname{Gauss}\left(\mu_{j}, \sigma_{j}\right)\right]+\frac{ N_{j}^{b}}{N_{j}}  \cdot \mathcal{P}_{j}^{b}\left({x}_{j} ; \vec{\lambda}_{j}^{b}\right),
		\label{equa:model}
	\end{equation}
	where $\mathcal{P}_{j}^{\mathrm{s}}$ is the signal shape PDF extracted from the signal MC samples, $N_j, N_j^s$ and $N_j^b$ are numbers of the total, signal, and background events, respectively;  ${x}_{j}$ is the variable corresponding to $M_{\mathrm {BC}}$;
%	 in Eq.~\eqref{equa:argus}; 
	 $\vec{\lambda}_{j}^{b}$ is a set of parameters in the function
	describing the background. The signal shape PDF is convolved with a Gaussian function describing the mass resolution difference between data and MC simulation. In this analysis, the background PDF  $ \mathcal{P}_{j}^{b}\left({x}_{j} ; \vec{\lambda}_{j}^{b}\right) $ is described by the ARGUS function~\cite{argus},
	\begin{equation}
%		f\left(M_{\rm BC}; E_{j}, c_j\right)=M_{\rm BC}\left[1-\left(\frac{M_{\rm B C}}{E_{				j}}\right)^{2}\right]^{0.5} \times e^{c_j \cdot\left[1-\left(\frac{M_{\rm B C}}{E_{j}}\right)^{2}\right]},
		f\left(x_j; E_{j}, c_j\right)=x_j\left[1-\left(\frac{x_j}{E_{	j}}\right)^{2}\right]^{0.5} \times e^{c_j \cdot\left[1-\left(\frac{x_j}{E_{j}}\right)^{2}\right]},
		\label{equa:argus}
	\end{equation}
	where $ E_{j}$ is the beam energy, and $c_j$ is the parameter of the ARGUS function. 
	The PDF of the reference decay  is similar to that of the signal decay.
	
	The yields of the signal decays are calculated by 
	\begin{equation}
		N_{j}^{\mathrm{s, sig}}=N_{j}^{\mathrm{s}, \mathrm{ref}} \cdot \frac{\mathcal{B}_{}^{\mathrm{inter, sig}}}{\mathcal{B}_{}^{\mathrm{inter}, \mathrm{ref}}} \cdot \frac{\epsilon_{j}^{\mathrm{sig}}}{\epsilon_{j}^{\mathrm{ref}}} \cdot R_{\mathrm{sig} / \mathrm{ref}},
	\end{equation}
	where $N_{j}^{\mathrm{s, sig}}$ is the yield of the signal decay, and 
	$N_{j}^{\mathrm{s,\mathrm{ref}}}$ is the yield of the reference decay;
	$\mathcal{B}^{\text {inter }}$ is the product of BF of the intermediate state, \textit{e.g.}, for the signal decay  $\Lambda_{c}^{+} \rightarrow \Sigma^{+} \eta$, $\mathcal{B}^{\text {inter }}=\mathcal{B}\left(\Sigma^{+} \rightarrow p \pi^{0}\right) \cdot \mathcal{B}\left(\pi^{0} \rightarrow \gamma \gamma\right) \cdot \mathcal{B}(\eta \rightarrow \gamma \gamma)$;
	$\epsilon^{\mathrm{sig}}_j$ and $\epsilon^{\mathrm{ref}}_j$ 
	are detection efficiencies for the signal and reference decays, respectively.
	$ R_{\mathrm{sig} / \mathrm{ref}}$ is the relative BF ratio. Different energy points share the same 
	value of $ R_{\mathrm{sig} / \mathrm{ref}}$.

	Overall, the parameters to derive for the signal decays are 
	\begin{equation}
		\begin{gathered}
			\vec{\lambda}_{j}=\left(N_{j}^{b}, \mu_{j}, \sigma_{j}, \vec{\lambda}_{j}^{b}, R_{\mathrm{sig} / \mathrm{ref}}\right) \\
		\end{gathered},
	\end{equation}
	while those for the reference decays are
	\begin{equation}
		\begin{gathered}
			\vec{\lambda}_{j}^{\mathrm{ref}}=\left(N_{j}^{b, \mathrm{ref}}, \mu_{j}, \sigma_{j}, \vec{\lambda}_{j}^{b, \mathrm{ref}}, N_{j}^{\mathrm{s}, \mathrm{ref}}\right)
		\end{gathered}.
	\end{equation}
	
	In this analysis, the $\mu_{j}$ and $\sigma_{j}$ are variable parameters of the Gaussian function,  $\mathcal{B}_{}^{\mathrm{inter, sig}}$ and $\mathcal{B}_{}^{\mathrm{inter}, \mathrm{ref}}$  are fixed to their values taken from the PDG~\cite{pdg2022}, and $ \epsilon_{j}^{\mathrm{ref}}$ and $\epsilon_{j}^{\mathrm{sig}} $ are fixed. The detection efficiencies, obtained by analyzing inclusive MC samples, are shown in  Table~\ref{table:efficiency}.
	
	\begin{table}[!htbp]
		\caption{The detection efficiencies~(in \%) at different energy points for each decay. The uncertainties are statistical.}
		\centering
		\resizebox{1.0\columnwidth}{!}{
			\begin{footnotesize}
				\setlength{\tabcolsep}{4mm}{
					\begin{tabular}{cccccc}
						\\
						\hline \hline $E_{\rm CMS}$~(GeV) & $\Lambda_{c}^{+} \rightarrow \Sigma^{+} \eta(\eta\rightarrow\gamma\gamma)$ & $\Lambda_{c}^{+} \rightarrow \Sigma^{+} \eta(\eta\rightarrow\pi^+\pi^-\pi^0)$ & $\Lambda_{c}^{+} \rightarrow \Sigma^{+} \pi^{0}$ & $\Lambda_{c}^{+} \rightarrow \Sigma^{+} \eta^{\prime}$ & $\Lambda_{c}^{+} \rightarrow \Sigma^{+} \omega$ \\
						\hline 
						4.600 &	$9.27\pm0.03$	& $4.44\pm0.03$	& $9.91\pm0.03$ 	& $3.14\pm0.02$	& $3.22\pm0.02$ \\
						4.612 &	$9.08\pm0.03$	& $4.18\pm0.03$	& $9.79\pm0.03$ 	& $2.84\pm0.02$	& $2.99\pm0.02$ \\
						4.628 &	$8.84\pm0.03$	& $4.07\pm0.03$	& $9.64\pm0.04$ 	& $2.83\pm0.02$	& $2.99\pm0.02$ \\
						4.641 &	$8.75\pm0.03$	& $3.98\pm0.03$	& $9.52\pm0.03$ 	& $2.83\pm0.02$	& $2.97\pm0.02$ \\
						4.661 &	$8.51\pm0.03$	& $3.90\pm0.03$	& $9.51\pm0.03$ 	& $2.80\pm0.02$	& $2.90\pm0.02$ \\
						4.682 &	$8.30\pm0.03$	& $3.75\pm0.03$	& $9.22\pm0.03$ 	& $2.80\pm0.02$	& $2.89\pm0.02$ \\
						4.699 &	$8.08\pm0.03$	& $3.70\pm0.03$	& $9.19\pm0.03$ 	& $2.79\pm0.02$	& $2.89\pm0.02$ \\
						\hline \hline
				\end{tabular}}
			\end{footnotesize}
		}
		\label{table:efficiency}
	\end{table}
	
\begin{figure}[!htbp]
		\centering
		\subfigure[$\Lambda_{c}^+ \rightarrow \Sigma^+ \eta( \eta\rightarrow\gamma \gamma) $.]{\includegraphics[width=0.3\textwidth]{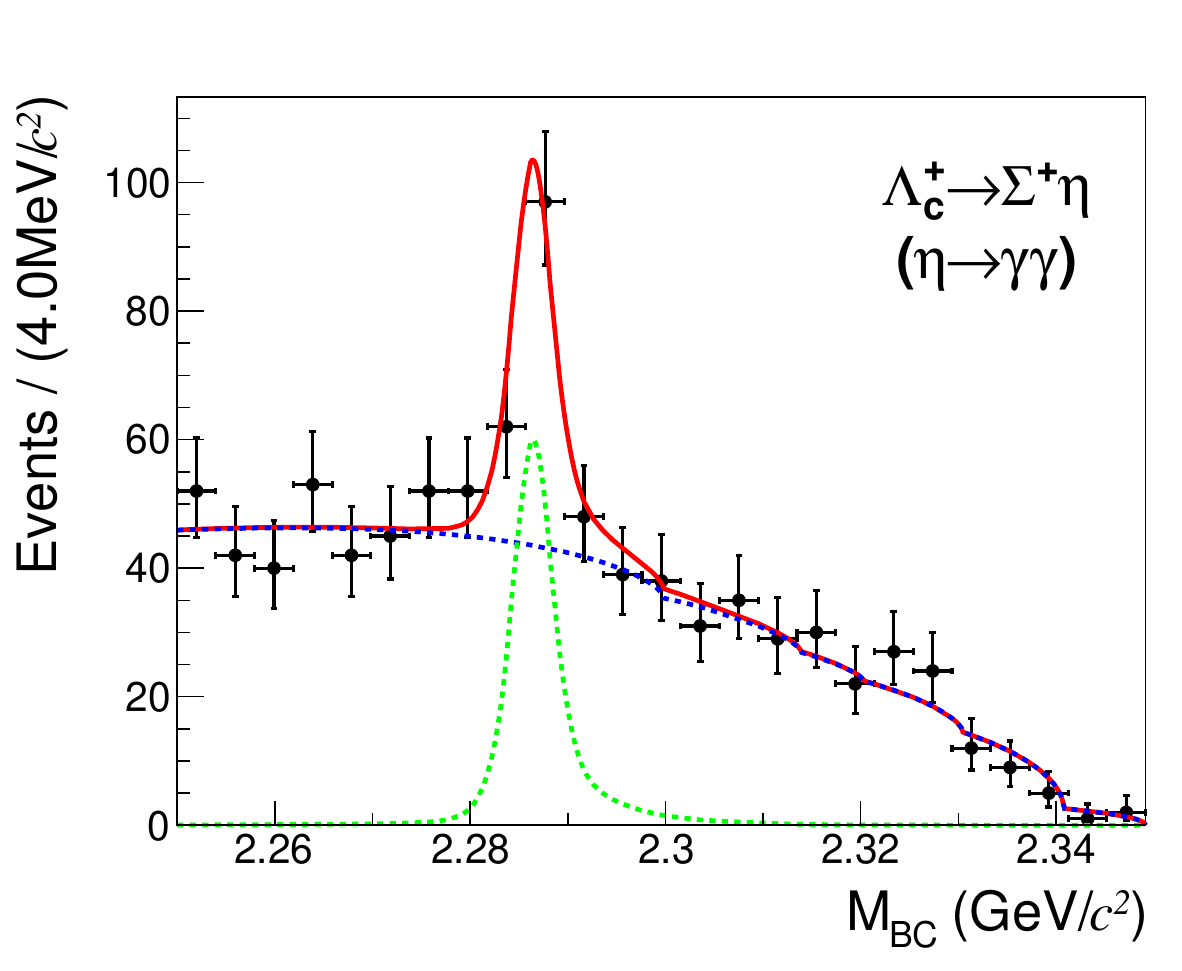}}
		\subfigure[$\Lambda_{c}^+ \rightarrow \Sigma^+ \eta(\eta\rightarrow\pi^+\pi^-\pi^0)$]{\includegraphics[width=0.3\textwidth]{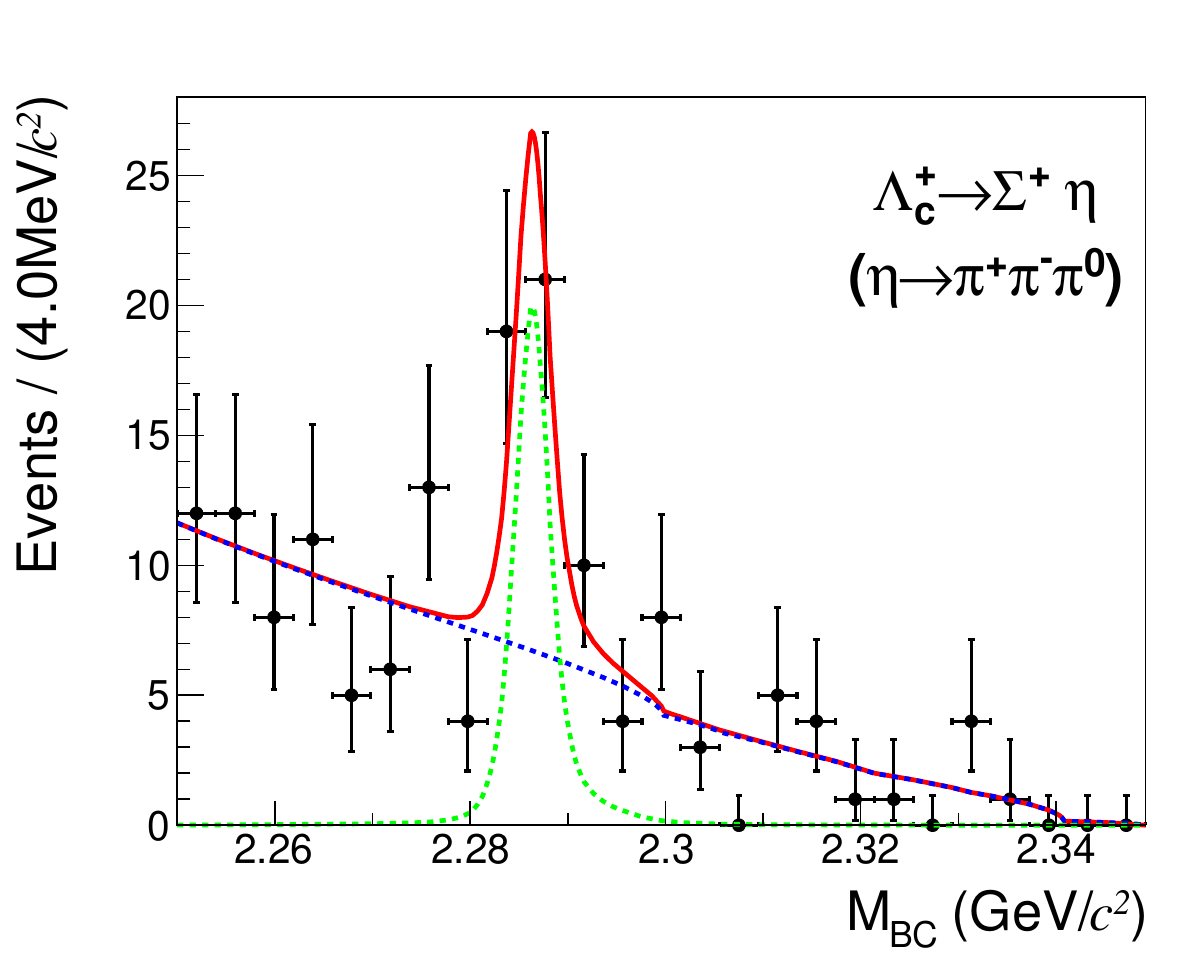}}
		\subfigure[$\Lambda_{c}^+ \rightarrow \Sigma^+ \eta'$]{\includegraphics[width=0.3\textwidth]{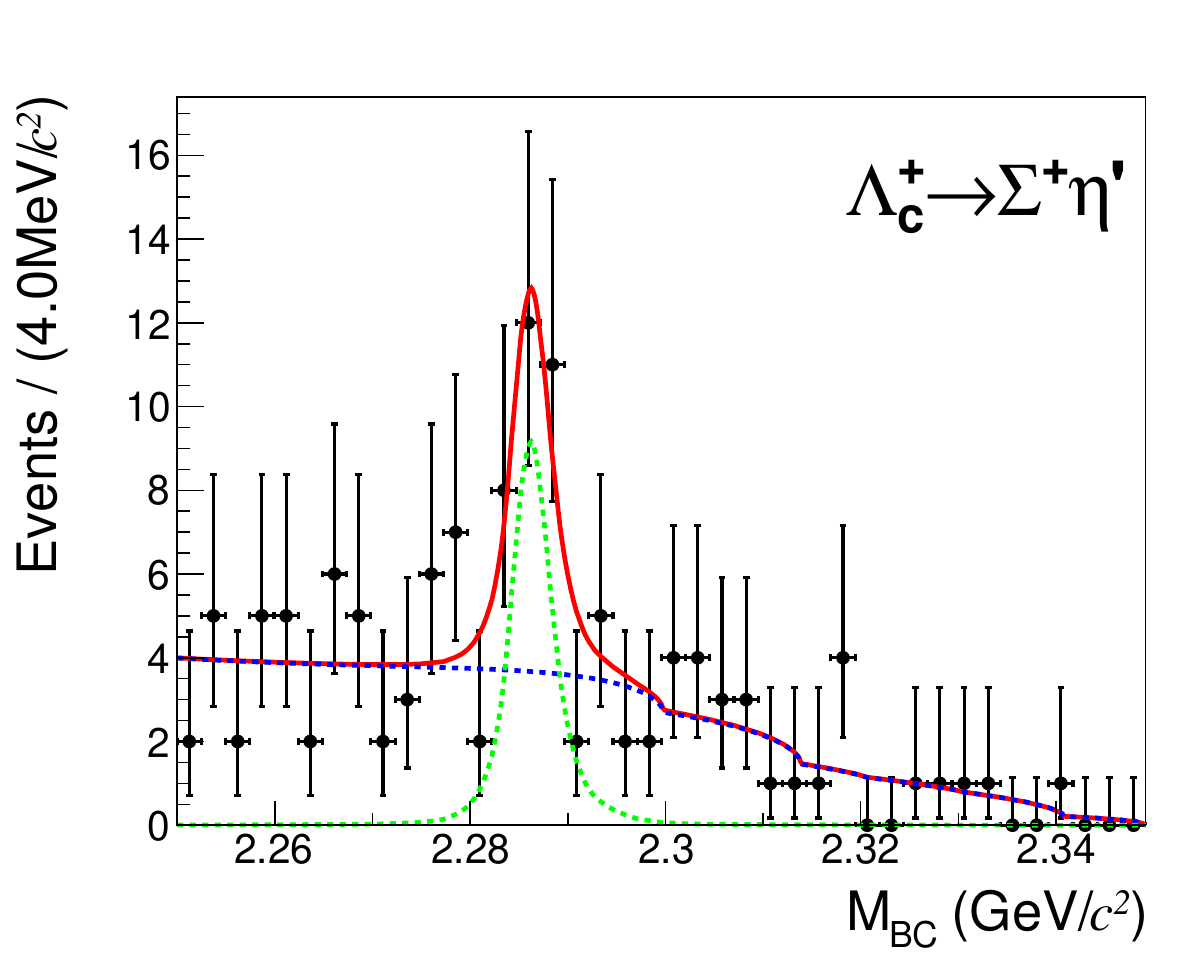}} \\
		\subfigure[$\Lambda_{c}^+ \rightarrow \Sigma^+ \pi^0 $]{\includegraphics[width=0.3\textwidth]{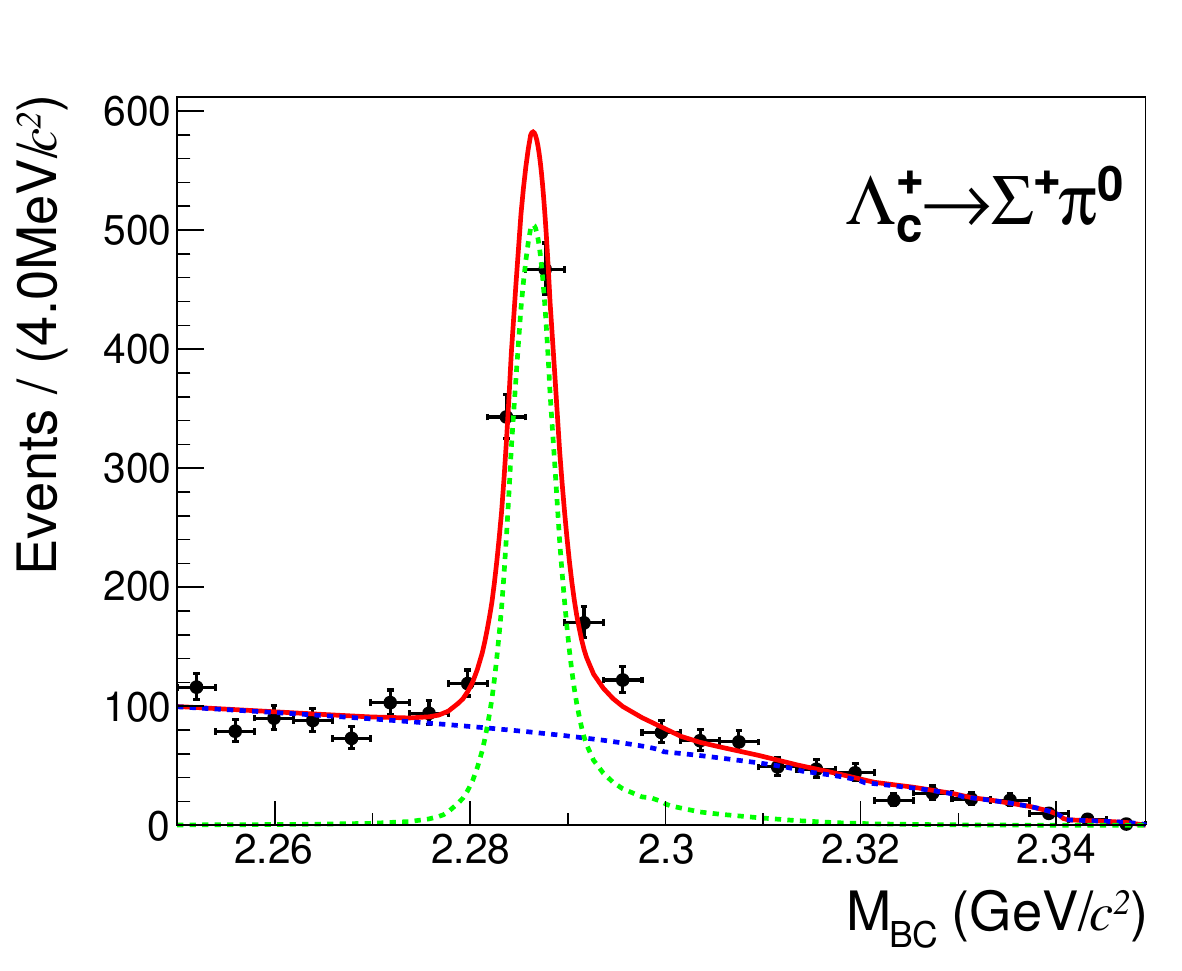}}
		\subfigure[$\Lambda_{c}^+ \rightarrow \Sigma^+ \omega$]{\includegraphics[width=0.3\textwidth]{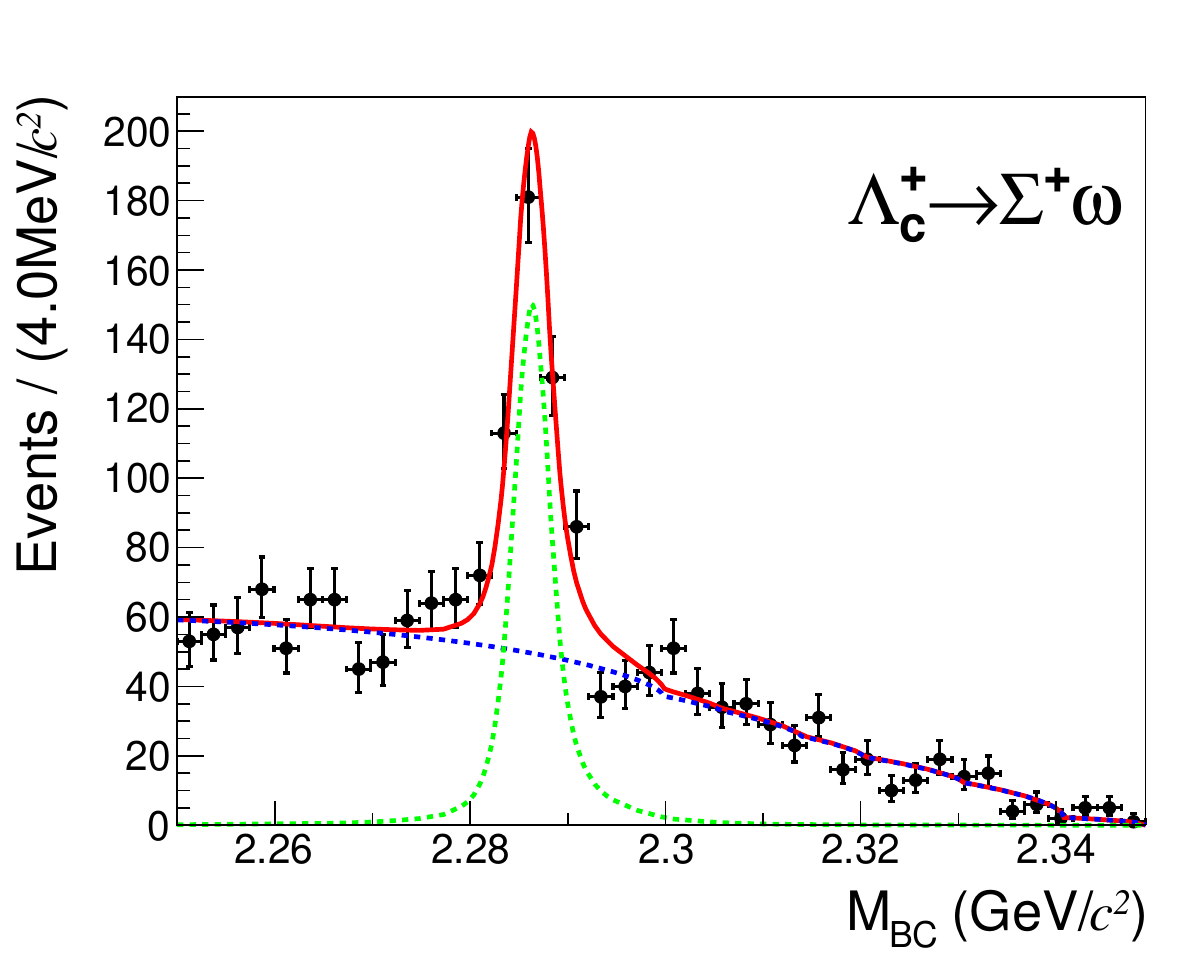}} \\	
		\caption{Projections of the simultaneous fit to the $M_{\rm BC}$ distributions at different energy points for each  decay mode. The black dots with error bars
			are data combining all energy points, the green lines show the signal, the dashed blue lines represent the background, and the red lines are the sum of the fit functions. }
		\label{fig:simultaneous_fit}
	\end{figure}

	We obtain the signal yields at each energy point by a simultaneous unbinned maximum likelihood fit to the $M_{BC}$ distributions, shown in Figure \ref{fig:simultaneous_fit}, using MIGRAD and HESSE in the RooFit~\cite{verkerke2003roofittoolkitdatamodeling} package; the results are summarized in Table \ref{table:fitresults}.
	For $\Lambda_{c}^{+} \rightarrow \Sigma^{+} \eta(\eta\rightarrow\gamma\gamma)$ and $\Lambda_{c}^{+} \rightarrow \Sigma^{+} \eta(\eta\rightarrow\pi^+\pi^-\pi^0)$, they are obtained by a simultaneous unbinned fit.
	The total yields of the signal decays $\Lambda_{c}^{+} \rightarrow \Sigma^{+} \eta$ and $\Lambda_{c}^{+} \rightarrow \Sigma^{+} \eta'$ are $121.6 \pm 19.5$  and $21.8 \pm 7.1$, respectively. 
	The  signal significance without considering the systematic uncertainties is $8.4\sigma$ for $\Lambda_{c}^{+} \rightarrow \Sigma^{+} \eta$  and $4.1\sigma$ for $\Lambda_{c}^{+} \rightarrow \Sigma^{+} \eta'$. 
	The BF of  $\Lambda_{c}^{+} \rightarrow \Sigma^{+} \eta$ relative to $\Lambda_{c}^{+} \rightarrow \Sigma^{+} \pi^0$  without considering the systematic uncertainties is $0.305 \pm 0.046 $, and the BF of  $\Lambda_{c}^{+} \rightarrow \Sigma^{+} \eta'$ relative to $\Lambda_{c}^{+} \rightarrow \Sigma^{+} \omega$  without considering the systematic uncertainties is $0.336 \pm 0.094$. 
	%%%%%%%%%%%%%%%%%%%%%%%%%%%%%%
	%%%%%%%%%%%%%%%%%%%%%%%%%%%%%%%
	\begin{table}[!htbp]
		\caption{Signal yields of each decay channel at different energy points, calculated by the BF ratios, which are the share parameters in the simultaneous fit.
			 The uncertainties are statistical.  }
		\centering
		\resizebox{1.0\columnwidth}{!}{
			\begin{footnotesize}
				\setlength{\tabcolsep}{2mm}{
					\sisetup{uncertainty-mode=separate,table-alignment-mode = format,
						table-number-alignment = center,}
					\begin{tabular}{cS[table-format=2.1(2),separate-uncertainty]
							S[table-format=3.1(2)]
							S[table-format=3.1(2)]
							S[table-format=3.1(2)]
							S[table-format=3.1(3)]}
						\\
						\hline \hline 
						{$E_{\mathrm{CMS}}$~(GeV)} & {$\Lambda_{c}^{+} \rightarrow \Sigma^{+} \eta'$} & {$\Lambda_{c}^{+} \rightarrow \Sigma^{+} \omega$} & {$\Lambda_{c}^{+} \rightarrow \Sigma^{+} \eta(\eta\rightarrow\gamma\gamma)$} & {$\Lambda_{c}^{+} \rightarrow \Sigma^{+} \eta(\eta\rightarrow\pi^+\pi^-\pi^0)$ } & {$\Lambda_{c}^{+} \rightarrow \Sigma^{+} \pi^{0}$ }\\
						\hline 
						4.600 & 2.7 \pm 0.9  & 43.2 \pm 8.3 & 14.5 \pm 2.8  & 4.0 \pm 0.8  &  127.2 \pm 13.4  \\
						4.612 & 0.2 \pm 0.2  & 3.5 \pm 3.0  & 2.3 \pm 0.7   & 0.6 \pm 0.2  & 20.3 \pm 5.3  \\
						4.628 & 2.1 \pm 0.8  & 35.3 \pm 7.8 & 11.9 \pm 2.4  & 3.2 \pm 0.6  &  106.9 \pm 12.6  \\
						4.641 & 3.1 \pm 1.0  & 50.9 \pm 9.0 & 11.2 \pm 2.3  & 2.9 \pm 0.6  &  100.2 \pm 12.5  \\
						4.661 & 2.7 \pm 0.9  & 44.6 \pm 8.2 & 12.5 \pm 2.5  & 3.3 \pm 0.7  &  114.5 \pm 13.3  \\
						4.682 & 8.1 \pm 2.4  & 131.3 \pm 14.7 & 34.2 \pm 6.1& 8.9 \pm 1.6  &  312.2 \pm 22.7  \\
						4.699 & 2.8 \pm 0.9  & 44.9 \pm 8.7 & 9.6 \pm 2.0  &  2.5 \pm 0.5  &  89.3 \pm 12.2  \\
						\hline\hline
					\end{tabular}
				}
			\end{footnotesize}
		}
		\label{table:fitresults}
	\end{table}

	\section{SYSTEMATIC UNCERTAINTY}
	\label{sec:systematic}
	Due to the limited statistics, the total uncertainties are dominated by the statistical errors. The systematic uncertainties associated with tracking and PID of charged pions, and photon selections are canceled in the measurement of the ratios of the  BFs. 
	Since the amount of simulated events is large, the statistical uncertainty due to the size of the MC samples is less than 0.1\%, which can be neglected.
	
	The following sources of systematic uncertainties in the measurement of the BFs are considered: the reconstruction of $\pi^{0}$ and $\eta$ states, the mass window for the  $\eta'$ and $\omega$ selections, the $p K^0_S$ veto, the $\chi^2 $ requirement for the best candidate, the modeling of the signal and of the background, the $M_{\rm BC}$  fit, the size of the MC samples and the BFs of the intermediate states.
%	 The uncertainties from tracking, PID and photon detection cancel out, since  both the signal and the reference decays have the same number of charged tracks and photons in their final states. 
	For the non-resonance contributions, we use a data-driven method to estimate this kind of background. 
	
	Table \ref{tab:sys_err} summarizes the sources of the systematic uncertainty in the BF measurements. The quadratic sum of all the uncertainties is taken as the total systematic uncertainty.
	The details of each item are as follows.
	%%%%%%%%%%%%%%%%%%%%%%%%%%%%%%%%%%%%%
	\begin{table*}[!hbp]
		\caption{Systematic uncertainties in percentage for the relative BF measurements. }
		\centering
		\footnotesize
		\begin{tabular}{ccc}
			\\
			\hline\hline Source & $\mathcal{B}\left(\Lambda_{c}^{+} \rightarrow \Sigma^{+} \eta\right)/\mathcal{B}\left(\Lambda_{c}^{+} \rightarrow \Sigma^{+} \pi^0\right)$ & $\mathcal{B}\left(\Lambda_{c}^{+} \rightarrow \Sigma^{+} \eta'\right)/\mathcal{B}\left(\Lambda_{c}^{+} \rightarrow \Sigma^{+} \omega\right)$ \\
			\hline
			$\eta$/$\pi^{0}$ reconstruction & $1.7$ & $0.8$ \\
			$\eta^{\prime}$ mass window & $-$ & $0.2$ \\
			$\omega$ mass window & $-$ & $0.2$ \\
			$p K^0_S$ veto & $0.2$ & $-$ \\
			$\chi^{2}$ requirement & $0.3$ & $0.3$ \\
			Signal model & $0.8$ & $1.1$ \\
			Non-resonant background & - & 10.7 \\
			$M_{\mathrm{BC}}$ fit & $0.4$ & $0.3$ \\
			Quoted BFs & $0.7$ & $1.5$ \\
			\hline Overall & $2.1$ & $10.9$ \\
			\hline\hline
		\end{tabular}
		\label{tab:sys_err}
	\end{table*}
	%%%%%%%%%%%%%%%%%%%%%%%%%%%%%%%%%%%%%
	
	\begin{itemize}
		\item \emph{$\pi^0$/$\eta$ reconstruction}.
		The systematic uncertainty for the $\pi^0$ reconstruction is studied with the control channel $D^0 \rightarrow K^- \pi^+ \pi^0$ in Ref~\cite{reco_pi0}. The control channels is studied to derive correction factors, which are then applied to the signal MC samples. 
		The  residual uncertainty of the correction factor is taken as the systematic uncertainty.
		For the systematic contribution from the $\eta$ reconstruction we proceeded in a similar way.
		The obtained systematic uncertainty is  $1.7\%$ for 
		$\frac{\mathcal{B}\left(\Lambda_{c}^{+} \rightarrow \Sigma^{+} \eta\right)}{\mathcal{B}\left(\Lambda_{c}^{+} \rightarrow \Sigma^{+} \pi^{0}\right)}$ 
		and $0.8 \%$ for $\frac{\mathcal{B}\left(\Lambda_{c}^{+} \rightarrow \Sigma^{+} \eta \prime\right)}{\mathcal{B}\left(\Lambda_{c}^{+} \rightarrow \Sigma^{+} \omega\right)}$.
		
		\item \emph{$\eta'$/$\omega$ mass windows}.	
		To estimate the systematic uncertainty caused by the $\eta' $ and $ \omega $ signal  mass windows, we smear their signal shapes by a Gaussian function with both the mean and width of 0.5 $\mathrm{MeV/}c^2$. The change of the nominal result is taken as the systematic uncertainty. Since there is no $\eta', \omega$ in the measurement of $\Lambda_{c}^{+} \rightarrow \Sigma^{+} \eta$, these part of systematic uncertainties are canceled.
		
		\item \emph {$p K_S^0$ veto}.
		For $\Lambda_{c}^{+} \rightarrow \Sigma^{+} \pi^0$, we reject combinations with $M(\pi^0\pi^0) \in  (0.440, 0.520) \; \mathrm{GeV/}c^2$ to veto the $\Lambda_c^+ \to p K_S^0$ background.
		To take into account the resolution difference between data and MC simulation, which may affect the detection efficiency, we change
		the veto range $M(\pi^0\pi^0)$ from $(0.440, 0.520)\; \mathrm{GeV/}c^2$ to $(0.450, 0.530)\; \mathrm{GeV/}c^2$ and $(0.430, 0.510)\; \mathrm{GeV/}c^2$, and estimate the related uncertainty as 0.2\%. In the measurement of $\Lambda_{c}^{+} \rightarrow \Sigma^{+} \eta'$, there is no $pK_S^0$ background,  this part of systematic uncertainty is canceled. 
		
		\item \emph{$\chi^2$ requirement}.
		Requirements on $\chi^2$ are applied to suppress background events, but some differences between data and MC simulation are present which can affect the measurement.
		We use control samples of  $\Lambda_{c}^{+} \rightarrow \Sigma^{+} \pi^0/\omega$ to smear the MC $\chi^2$ distributions to better reproduce the data, and we recalculate the BF ratios. The difference between the new values and the original ones is considered as the systematic uncertainty due to the $\chi^2$ requirement.
		
		\item \emph{Signal model}.
		The estimation of the efficiency values depends on the assumption about the helicity angle. In charmed baryon decays, the helicity angular distribution  is determined by the asymmetry parameter $\alpha$ varied by $\pm 1\sigma$ from PDG\cite{pdg2022}. 
		The systematic uncertainties from the signal model of the reference decays $\Lambda_{c}^+ \rightarrow \Sigma^+ \pi^0 $ and  $\Lambda_{c}^+ \rightarrow \Sigma^+ \omega$  are negligible.
		The uncertainty of the signal model is $0.8\%$ for  $\Lambda_{c}^+ \rightarrow \Sigma^+ \eta$
		and $1.1\%$ for $\Lambda_{c}^+ \rightarrow \Sigma^+ \eta'$.

		\item \emph{Non-resonant background}.
		The non-resonant $\Lambda_c^+ \rightarrow p \pi^0 \eta/\eta'$ decays are both Cabibbo-suppressed, and their BFs are about one order of magnitude smaller than those for the two signal decays. Besides, we check the $\Sigma^+$ sideband regions, no peak is found.
		Therefore, the non-$\Sigma^+$ contributions are negligible. 
		For the $\Lambda_c^+ \rightarrow \Sigma^+ \pi^+ \pi^- \pi^0/\eta$ decays, we use a data-driven method to estimate their contribution.
		We examine the $M_{\rm{BC}}$ distribution of the events in the $\eta'/\omega$  sideband region searching for a peak in the $M_{\rm{BC}}$ signal region.  Since no $\eta'$ peak is found, its related uncertainty is safely neglected.

		To consider the non-resonant background for the $\Lambda_c^+ \rightarrow \Sigma^+ \omega$ channel, we perform a simultaneous fit on the signal  and the reference decays for events in the $\omega$  signal and sideband regions.
		
		For the $\Lambda_c^+ \rightarrow \Sigma^+ \eta'$ channel, the fit model is the same as that in the $\Lambda_c^+ \rightarrow \Sigma^+ \eta$ analysis shown in Eq.~\eqref{equa:model}, and the fit result is shown in Figure \ref{fig:simufit_sideband}(a). For  $\Lambda_c^+ \rightarrow \Sigma^+ \omega$, the fit model in the $\omega$ signal region is constructed as 
		\begin{equation}
			\mathcal{P}_{j}\left(\vec{x}_{j} ; \vec{\lambda}_{j}\right)=\frac{N_{j}^{\mathrm{s}}}{N_{j}} \cdot\left[\mathcal{P}_{j}^{\mathrm{s}} \otimes \operatorname{Gauss}\left(\mu_{j}, \sigma_{j}\right)\right]+\frac{ N_{j}^{b1}}{N_{j}}  \cdot \mathcal{P}_{j}^{b1}\left(\vec{x}_{j} ; \vec{\lambda}_{j}^{b}\right) + 
			\frac{ N_{j}^{b2}}{N_{j}}  \cdot\left[\mathcal{P}_{j}^{{b2}} \otimes \operatorname{Gauss}\left(\mu_{j}, \sigma_{j}\right)\right],
			\label{equa:model_ref}
		\end{equation}
		where $ \mathcal{P}_{j}^{b1}\left(\vec{x}_{j} ; \vec{\lambda}_{j}^{b1}\right) $ is described by the ARGUS function~\cite{argus} in the $\omega$  signal region, and $ \mathcal{P}_{j}^{{b2}} \otimes \operatorname{Gauss}\left(\mu_{j}, \sigma_{j}\right) $ is to describe the non-resonant background. $N_{j}^{b1}$ and $N_{j}^{b2}$ are the ARGUS and the non-resonant background yields, respectively. The fit result is shown in Figure \ref{fig:simufit_sideband}(b). For  $\Lambda_c^+ \rightarrow \Sigma^+ \omega$, the fit model in the $\omega$ sideband region is 
		\begin{equation}
			\mathcal{P}_{j}\left(\vec{x}_{j} ; \vec{\lambda}_{j}\right)=
			\frac{ N_{j}^{b3}}{N_{j}}  \cdot \mathcal{P}_{j}^{b1}\left(\vec{x}_{j} ; \vec{\lambda}_{j}^{b}\right) + 
			{{\frac{ N_{j}^{b2}/r_1}{N_{j}} }}
			\cdot\left[\mathcal{P}_{j}^{{b2}} \otimes \operatorname{Gauss}\left(\mu_{j}, \sigma_{j}\right)\right],
			\label{equa:model_refsb}
		\end{equation}
		where $N_{j}^{b3}$ is the ARGUS background yield in the $\omega$ sideband region, $r_1$ is the ratio between the integral of the signal PDF in the $\omega$  signal region and that of the background PDF in the $\omega$  sideband regions in the fit to the $\pi^+\pi^-\pi^0$ invariant mass distribution.
		The result of the simultaneous fit is shown in Figure \ref{fig:simufit_sideband}(c).
		Including all non-resonant contributions, this systematic uncertainty is assigned as  10.7\%.

		\begin{figure}[!htbp]
			\centering
			\subfigure[] {\includegraphics[width=0.325\textwidth]{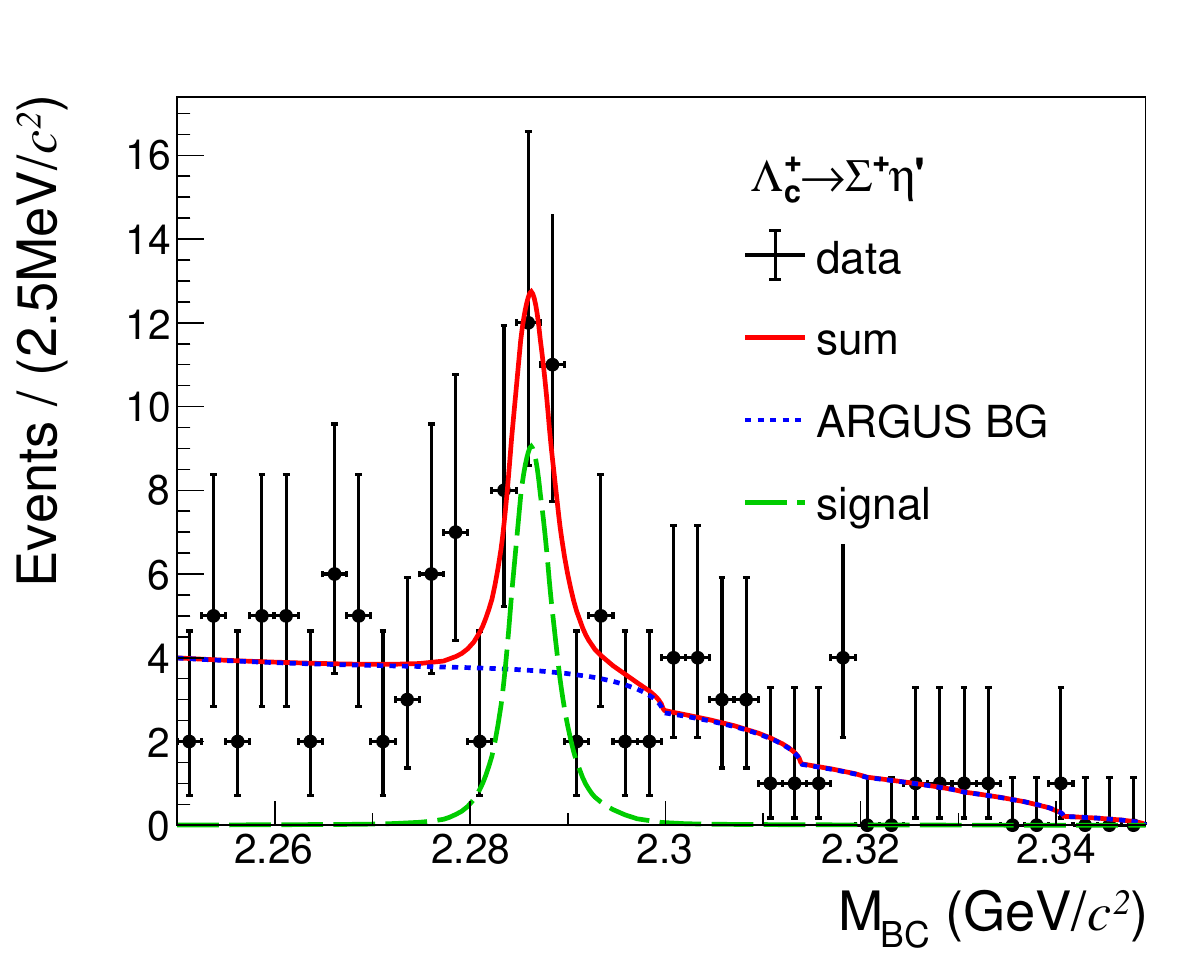}}\
			\subfigure[] {\includegraphics[width=0.325\textwidth]{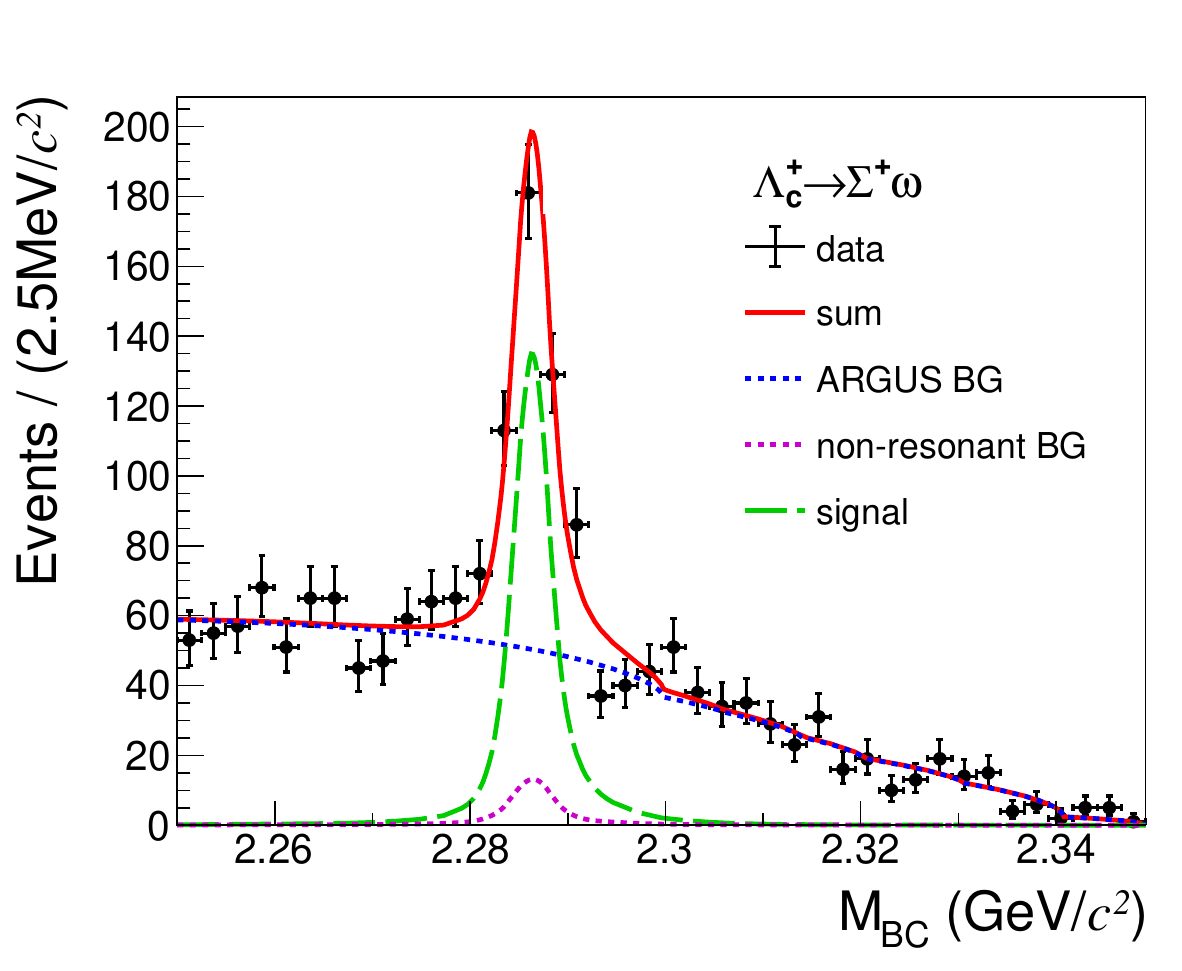}}\
			\subfigure[] {\includegraphics[width=0.325\textwidth]{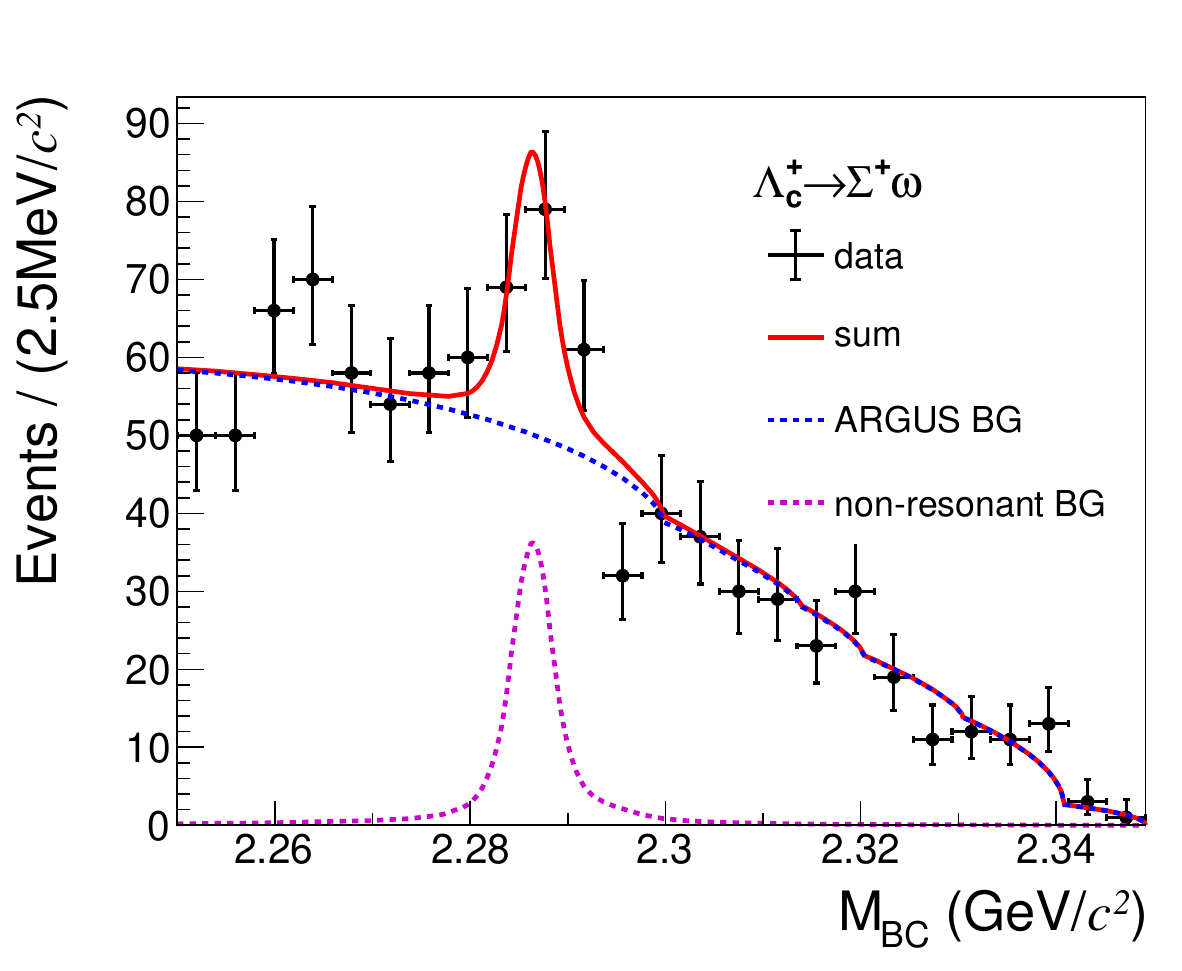}}\\
			\caption{Simultaneous fit to the ${M_{\rm BC}}$ distributions. (a) and (b) correspond to  the $\omega$ signal region, (c) corresponds to the $\eta'/\omega$  sideband region. The black dots with error bars
				are data combined from all energy points, the dashed green lines are the signal, the dotted blue and magenta lines are the backgrounds, and the continuous red lines are the sum of the fit functions. }
			\label{fig:simufit_sideband}
		\end{figure}

		\item \emph{${M}_{\rm BC}$ fit}.
		The systematic uncertainty  in the $M_{\rm BC}$ fit is mainly due to the ARGUS functions and to the background description. The relevant systematic uncertainty is estimated by using  an alternative background shape, by changing  the high-end cutoff of the ARGUS function by $\pm 0.005$ GeV/$c^2$. The largest BF change is assigned as the systematic uncertainty.
		
		\item \emph {Quoted BFs}.
		In the nominal analysis, the branching fractions of the $\eta' \to \pi^+ \pi^- \eta$, $\omega \to \pi^+ \pi^- \pi^0$, $\Sigma^+ \to p \pi^0$, and $\pi^0/\eta \to \gamma \gamma$ decays are quoted from the PDG~\cite{pdg2022},  and their uncertainties are taken into account. 
		The uncertainties
		from $\mathcal{B}(\Sigma^+ \rightarrow p \pi^0)$ cancel out. The total uncertainty is 0.7\% for $\frac{\mathcal{B}\left(\Lambda_{c}^{+} \rightarrow \Sigma^{+} \eta\right)}{\mathcal{B}\left(\Lambda_{c}^{+} \rightarrow \Sigma^{+} \pi^{0}\right)}$ 
		and 1.5\% for  $\frac{\mathcal{B}\left(\Lambda_{c}^{+} \rightarrow \Sigma^{+} \eta \prime\right)}{\mathcal{B}\left(\Lambda_{c}^{+} \rightarrow \Sigma^{+} \omega\right)}$.

	\end{itemize}

	\section{SUMMARY}
	\label{sec:summary}
	\hspace{1.5em}
	By analyzing  $e^+e^-$ collision data taken at center-of-mass energies $\sqrt{s}$ between  4.600 and 4.699 $~\gev$ with the BESIII detector corresponding to an integrated luminosity of $4.5~\rm fb^{-1}$, and using a single-tag reconstruction method,
	we have measured the BF of $\Lambda_{c}^+ \rightarrow \Sigma^+ \eta$ relative to $\Lambda_{c}^+ \rightarrow \Sigma^+ \pi^0$ as $0.305 \pm 0.046_{\rm stat.} \pm 0.007_{\rm syst.}$,
	and the BF of $\Lambda_{c}^+ \rightarrow \Sigma^+ \eta'$ relative to $\Lambda_{c}^+ \rightarrow \Sigma^+ \omega$ as $0.336 \pm 0.094_{\rm stat.} \pm 0.037_{\rm syst.}$.  
	
	Table~\ref{tab:zhou_me} shows the comparison of the BF ratios with those based on data taken at $\sqrt{s} = 4.6~\gev$  by the BESIII detector~\cite{BESIII:2018cdl} corresponding to an integrated luminosity of $\rm 567~pb^{-1}$. They are compatible within $1\sigma$  for ${\mathcal{B}\left(\Lambda_{c}^{+} \rightarrow \Sigma^{+} \eta\right)}/{\mathcal{B}\left(\Lambda_{c}^{+} \rightarrow \Sigma^{+} \pi^{0}\right)}$ and within $2\sigma$  for ${\mathcal{B}\left(\Lambda_{c}^{+} \rightarrow \Sigma^{+} \eta^{\prime}\right)}/{\mathcal{B}\left(\Lambda_{c}^{+} \rightarrow \Sigma^{+} \omega\right)}$. This consistency is based on the larger uncertainty of the BESIII previous measurement. 
	Benefiting from the larger collected data sample, we find evidence for the decay $\Lambda_c^+ \rightarrow \Sigma^+ \eta$, and the total uncertainties of the measured  BF ratios improve by about two-thirds for both decays. 
	
	Table~\ref{tab:summary_exp}  shows a comparison of our BFs of $\Lambda_{c}^{+} \rightarrow \Sigma^{+} \eta$ and $\Lambda_{c}^{+} \rightarrow \Sigma^{+} \eta'$ with the Belle measurements~\cite{2023Belle}, where the uncertainties are statistical, systematic, and from $\mathcal{B}\left(\Lambda_{c}^{+} \rightarrow \Sigma^{+} \pi^0\right) $ or $\mathcal{B}\left(\Lambda_{c}^{+} \rightarrow \Sigma^{+} \omega\right) $ obtained from PDG~\cite{pdg2022}.
	They are compatible within $1\sigma $  for both decays.  
	The improved precision of our results clarifies the difference between the Belle values and the previous BESIII measurements. 
	
	The comparison of the BF predictions under different theoretical models is shown in Table.~\ref{tab:summary_theory}. 
	The $\mathcal{B}\left(\Lambda_{c}^{+} \rightarrow \Sigma^{+} \eta\right)$ value presented in this work is compatible with Zhao~\cite{2020Zhao}, Geng~\cite{2019Geng}, and Cheng's~\cite{Cheng_2025oyr} works within $1\sigma$ based on $SU(3)$ flavor symmetry. 
	For  the $\mathcal{B}\left(\Lambda_{c}^{+} \rightarrow \Sigma^{+} \eta'\right)$ value, 
	our measurement is consistent with Cheng's result \cite{Cheng_2025oyr} at the $1\sigma$ level and agrees with the preditions reported by Geng \cite{2019Geng} and Zhao \cite{2020Zhao} within $2\sigma$.
	% the result reported by Cheng~\cite{Cheng_2025oyr} is consistent within $1\sigma$, whereas those presented by Geng~\cite{2019Geng} and Zhao~\cite{2020Zhao} agree within $2\sigma$.
	
	Furthermore, we determine the BF ratio 
	$\frac{\mathcal{B}\left(\Lambda_{c}^{+} \rightarrow \Sigma^{+} \eta'\right)}{\mathcal{B}\left(\Lambda_{c}^{+} \rightarrow \Sigma^{+} \eta\right)}
	=1.73 \pm 0.22_{\rm stat.} \pm 0.16_{\rm syst.}$,
%	 based on the results in Table~\ref{tab:summary_exp}.
%	This value is consistent with the result reported by Cheng~\cite{Cheng_2025oyr} and in agreement with previous measurements from BESIII and Belle~\cite{2023Belle,BESIII:2018cdl}.
	% which is consistent with Cheng~\cite{Cheng_2025oyr}, and also 
	which is consistent with the previous measurements at BESIII and Belle~\cite{BESIII:2018cdl,2023Belle}. 
	The obtained results will help understand the decay mechanisms of charmed baryon decays.
	
	%%%%%%%%%%%%%%%%%%%%%%%%%%%%%%%%%%
	\begin{table}
		\caption{Comparison of the determined BF ratios with the previous BESIII measurements.}
		\centering
		\footnotesize
		\begin{tabular}{lcc}
			\hline  BF ratio & BESIII~\cite{BESIII:2018cdl}& This work\\
			\hline ${\mathcal{B}\left(\Lambda_{c}^{+} \rightarrow \Sigma^{+} \eta\right)}/{\mathcal{B}\left(\Lambda_{c}^{+} \rightarrow \Sigma^{+} \pi^{0}\right)}$ 
			& $0.35 \pm 0.16 \pm 0.02 $ & $0.305 \pm 0.046 \pm 0.007 $ \\
			${\mathcal{B}\left(\Lambda_{c}^{+} \rightarrow \Sigma^{+} \eta^{\prime}\right)}/{\mathcal{B}\left(\Lambda_{c}^{+} \rightarrow \Sigma^{+} \omega\right)}$ 
			&  $0.86\pm 0.34 \pm 0.04 $ & $0.336\pm0.094 \pm 0.037 $\\
			\hline
		\end{tabular}
		\label{tab:zhou_me}
	\end{table}
	%%%%%%%%%%%%%%%%%%%%%%%%%%%%%%%%%%
	\begin{table}
		\caption{Comparison of the obtained BFs with other experimental measurements and the world average values (in units of \%).}
		\centering
		\footnotesize
		\resizebox{1.\columnwidth}{!}{
			\begin{tabular}{lccccccc}
				\hline Decay     &  PDG~\cite{pdg2022} & CLEO~\cite{1995RAmmar} 
				& Belle~\cite{2023Belle} 
				& BESIII\cite{BESIII:2018cdl}  &This work\\
				\hline$\Lambda_{c}^{+} \rightarrow \Sigma^{+} \eta$ 
				& $0.44 \pm 0.2 $ & $0.70 \pm 0.23 $ 
				
				& $0.314\pm0.035\pm0.011\pm0.025$ 	& $0.41 \pm 0.20 $  & $0.381\pm0.058\pm0.009\pm0.027$ \\
				$\Lambda_{c}^{+} \rightarrow \Sigma^{+} \eta'$  & $1.5\pm0.6$ & -- 
				& $0.416\pm0.075\pm0.021\pm0.033$ & $1.34\pm0.56 $ & 
				$0.571\pm0.160\pm0.063\pm0.067$\\
				\hline
			\end{tabular}
		}
		\label{tab:summary_exp}
	\end{table}
	%%%%%%%%%%%%%%%%%%%%%%%%%%%%%%%%%%
	\begin{table}
		\caption{Comparison of the BF predictions under different theoretical models (in units of \%).}
		\centering
		\footnotesize
		\resizebox{\textwidth}{!}{
			\begin{tabular}{lccccccccc}
				\hline Decay mode & Uppal~\cite{1994TUppal}  & Körner\cite{1992Korner} & Sharma\cite{1999Sharma} & Zenczykowski\cite{1994Zenczykowski}   & Ivanov\cite{1998Ivanov} & Zou\cite{2020Zou} & Geng\cite{2019Geng} & Zhao\cite{2020Zhao} & Cheng\cite{Cheng_2025oyr}\\
				\hline$\Lambda_{c}^{+} \rightarrow \Sigma^{+} \eta$ & 0.22 & 0.16 & $0.57$ & $0.90$ 
				& 0.11 & 0.74 & $0.32 \pm 0.13 $ & $0.47\pm0.22$ & $0.35\pm0.04$\\
				$\Lambda_{c}^{+} \rightarrow \Sigma^{+} \eta'$ & 0.05 &1.28 & $0.10$ & $0.11$ & 0.12 & -- & $1.44\pm0.56 $& $0.93\pm0.28$  & $0.40\pm0.07$\\
				\hline
		\end{tabular}}
		\label{tab:summary_theory}
	\end{table}
	%%%%%%%%%%%%%%%%%%%%%%%%%%%%%%%%%%
	%%%%%%%%%%%%%%%%%%%%%%%%%%%%%%%%%%%%%%%%%%%%%%
	%%%%%%%%%%%%%%%%%%%%%%%%%%%%%%%%%%%%%%%%%%%%%%

	\acknowledgments
	The BESIII Collaboration thanks the staff of BEPCII and the IHEP computing center for their strong support. This work is supported in part by National Key R\&D Program of China under Contracts Nos. 2023YFA1606000, 2020YFA0406300, 2020YFA0406400; National Natural Science Foundation of China (NSFC) under Contracts Nos. 11635010, 11735014, 11935015, 11935016, 11935018, 12025502, 12035009, 12035013, 12061131003, 12192260, 12192261, 12192262, 12192263, 12192264, 12192265, 12221005, 12225509, 12235017, 12361141819; the Chinese Academy of Sciences (CAS) Large-Scale Scientific Facility Program; the CAS Center for Excellence in Particle Physics (CCEPP); Joint Large-Scale Scientific Facility Funds of the NSFC and CAS under Contract No. U1832207; 100 Talents Program of CAS; The Institute of Nuclear and Particle Physics (INPAC) and Shanghai Key Laboratory for Particle Physics and Cosmology; German Research Foundation DFG under Contracts Nos. 455635585, FOR5327, GRK 2149; Istituto Nazionale di Fisica Nucleare, Italy; Ministry of Development of Turkey under Contract No. DPT2006K-120470; National Research Foundation of Korea under Contract No. NRF-2022R1A2C1092335; National Science and Technology fund of Mongolia; National Science Research and Innovation Fund (NSRF) via the Program Management Unit for Human Resources \& Institutional Development, Research and Innovation of Thailand under Contract No. B16F640076; Polish National Science Centre under Contract No. 2019/35/O/ST2/02907; The Swedish Research Council; U. S. Department of Energy under Contract No. DE-FG02-05ER41374

	\bibliographystyle{JHEP}   
	\bibliography{ref.bib} 
	
	\newpage
	\noindent
%	\collaboration{
%	{\large The BESIII collaboration}\\
	\\
	%% Saved at => 2024-03-08
M.~Ablikim$^{1}$\orcidlink{0000-0002-3935-619X},
M.~N.~Achasov$^{4,c}$\orcidlink{0000-0002-9400-8622},
P.~Adlarson$^{76}$\orcidlink{0000-0001-6280-3851},
O.~Afedulidis$^{3}$\orcidlink{0009-0006-2899-9946},
X.~C.~Ai$^{81}$\orcidlink{0000-0003-3856-2415},
R.~Aliberti$^{35}$\orcidlink{0000-0003-3500-4012},
A.~Amoroso$^{75A,75C}$\orcidlink{0000-0002-3095-8610},
Q.~An$^{58,72,a}$,
Y.~Bai$^{57}$\orcidlink{0000-0001-6593-5665},
O.~Bakina$^{36}$\orcidlink{0009-0005-0719-7461}
% ...
\\
\vspace{0.2cm}
(BESIII Collaboration)\\
\vspace{0.2cm} {\it
$^{1}$ Institute of High Energy Physics, Beijing 100049, People's Republic of China\\
% ...
$^{3}$ Bochum  Ruhr-University, D-44780 Bochum, Germany\\
$^{4}$ Budker Institute of Nuclear Physics SB RAS (BINP), Novosibirsk 630090, Russia\\
% ...
$^{35}$ Johannes Gutenberg University of Mainz, Johann-Joachim-Becher-Weg 45, D-55099 Mainz, Germany\\
$^{36}$ Joint Institute for Nuclear Research, 141980 Dubna, Moscow region, Russia\\
% ...
$^{57}$ Southeast University, Nanjing 211100, People's Republic of China\\
$^{58}$ State Key Laboratory of Particle Detection and Electronics, Beijing 100049, Hefei 230026, People's Republic of China\\
% ...
$^{72}$ University of Science and Technology of China, Hefei 230026, People's Republic of China\\
% ...
$^{75}$ University of Turin and INFN, (A)University of Turin, I-10125, Turin, Italy; (B)University of Eastern Piedmont, I-15121, Alessandria, Italy; (C)INFN, I-10125, Turin, Italy\\
$^{76}$ Uppsala University, Box 516, SE-75120 Uppsala, Sweden\\
% ...
$^{81}$ Zhengzhou University, Zhengzhou 450001, People's Republic of China\\

\vspace{0.2cm}
$^{a}$ Deceased\\
% ...
$^{c}$ Also at the Novosibirsk State University, Novosibirsk, 630090, Russia\\
% ...

}
%% ends here %%

	%\end{linenumbers}
\end{document}